\documentclass[11pt]{article}
\usepackage{amsmath}
\usepackage{accents}  
\usepackage{amssymb}  
\newcommand{\ft}[2]{{\textstyle\frac{#1}{#2}}}
\usepackage{graphicx}
\begin{document}

\title{ Three dimensional black strings: instabilities and asymptotic charges}

\author{Ph. Spindel\footnote{{\texttt Email : philippe.spindel@umons.ac.be}}\\Service de Physique de l'Univers, Champs et Gravitation,\\ Universit\'e de Mons,\\ Facult\'e des Sciences,20, Place du Parc, B-7000 Mons, Belgium\\
\strut\\
Service de Physique Th\'eorique,
Universit\'e Libre de Bruxelles\\
Bld du Triomphe CP225, 1050 Brussels, Belgium}
\vspace{10pt}
\maketitle
\begin{abstract}

Three-dimensional Einstein gravity coupled to zero, one and two forms is solved in terms of a polyhomogeneous asymptotic expansion, generalising stationary black string solutions.  From first order terms we obtain, in closed form, a new solution evolving from the stationary black string structure to a geometry developing a singularity in the future. This new solution itself may be extended to more general ones. Taking into account subleading terms of the asymptotic expansion, both singularities in the past and the future occur. This demonstrates the unstable character of the stationary black string : tiny perturbations generated by terms  breaking the rotational invariance of the stationary black string configurations lead to cosmological-like singularities.     The symmetry algebra of the conserved charges also is determined: it is a finite dimensional one. In the general case the surface charge associated to energy is not integrable. However we identify a sub-class of solutions, admitting asymptotic symplectic symmetries and as a consequence conserved charges that appear to be integrable. 
\end{abstract}

\section{Introduction}
  Black holes are some of the most fascinating solutions of general relativity. They verify thermodynamic relations that provide a bridge between the classical theory of gravity and its quantized version.  Three dimensional black strings appear naturally as target space geometries of gauged WZW models \cite{Witten:1991yr,Horne:1991gn,Detournay:2005fz,Detournay:2018cbf} {\it i.e. }of marginal deformation of the $SL(2,\mathbb R)$ {\sc WZW} model. They provide solutions of the field equations deduced from a low energy string Lagrangian describing the coupling of Einstein gravity to a dilaton (a scalar resulting from quantum effects), a Maxwellian (abelian one form) and a Kalb--Ramond (two form) field.

Here we present a large class of   solutions of 3 dimensional relativity inspired from heterotic string theory. The interest
of such solutions is that they unable us to understand some aspects of quantum gravity, at least in the framework of low dimensional gravity. For instance they suggest interesting deformations of the underlying sigma-model, as the one presented in section {\bf 5.2} of ref.\cite{Detournay:2018cbf}, inspired from the new solutions presented hereafter (Eqs[\ref{fMS}--\ref{HMS}]). They also offer a framework to study asymptotic dynamics in a much more easier way that in four dimensions. Indeed it is well known that generic solutions of Einstein equations dont admit any local symmetry (Killing vector field) that allows to define exact conserved quantities. However, since the seminal work of Arnowitt, Deser and Misner \cite{Arnowitt:1962hi} (see also refs\cite{Abbott:1981ff,Regge:1974zd}),  the existence and relevance of asymptotic symmetries has been put into evidence. In favorable case these asymptotic symmetries lead to asymptotic conservation laws whose associated charges constitute the generators of a symmetry algebra underlying the asymptotic dynamic.

An important part of this work is devoted to the integration of Einstein equations coupled to a dilaton (scalar), a Maxwellian (abelian one-form) and a Kalb-Ramond (two form) fields. This dynamical system offers, at least,  two distinct covariant phase space sectors exhibiting black holes.  The first one, where the r\^ole of the cosmological constant is played by the Kalb-Ramond  flux, leads to the well known {\sc btz} black holes configurations \cite{Banados:1992wn}. The second one, put into light by Horne and Horowitz \cite{Horne:1991gn}, also provides  black string/brane configurations but where the (singular) behaviour of the dilaton field dictates the behaviour of the geometry. As we shall show, the latter defines a  covariant phase space where this (two degrees of freedom) dynamical problem appears to be exactly integrable in an asymptotic expansion scheme involving inverse powers  and logarithms of a radial coordinate. 

These three dimensional configurations constitute toy models for studying several aspects of gravity. In particular, we derive closed forms of exact solutions exhibiting dynamical  configurations.
This was a totally unexpected result. We have also obtained a large class of solutions, extending the stationary Horne and Horowitz black string solution, to much more general configurations, without any symmetry. However they all appear to be unstable    in the sense that under a tiny perturbation breaking the rotational invariance, the static black string configuration develops new cosmological-like curvature singularities. This result answers to a conjecture stated in ref. \cite{Horne:1991gn}, conjecture based on the first order analysis   developed in ref. \cite{ChHa}.

We have integrated the field equations similarly  to the approach of ref. \cite{Barnich:2015jua} in the three-dimensional framework of Maxwell-Einstein equations. However let us already mention that some non-trivial differences with this work will emerge along the thread of our analysis. Indeed the presence of a dilaton, singular at spatial infinity,  leads to a modification of the asymptotic behaviour of  other fields. It is the dilaton itself,  multiplicatively coupled to the abelian field, that will provide the expected logarithmic terms needed to obtain electric charges in three spacetime dimensions. They are  extracted  from sub-leading terms appearing during the expansion of the gauge fields. Of course the dilaton multiplying the abelian and the Kalb-Ramond can be reabsorbed in their definitions, but at the price of introducing (apparently) more complicated coupling terms.

A second part  of this work is devoted to constructing configurations
 leading to finite asymptotic conserved charges.  
  In particular we obtain that subdominant terms of the asymptotic expansion of the scalar fields occur as non integrable components in the expression of the charge associated to energy [see eq. (\ref{genEldE})]. Our considerations are based on a lemma [proved in appendix (\ref{AppSurfCh})] which guarantees the covariance of the expressions of the charge density we consider. This  allows us to perform the calculation in a gauge where the expressions for the various field components are simpler and from them infer conditions that lead to finite charges in more general situations,  obtained by acting with so-called "large diffeomorphisms" [a point also discussed in appendix (\ref{AppSurfCh})].

\section{Covariant phase space configurations}
The geometries we consider in this work are solutions of the field equations provided by the diffeomorphism invariant action:
\begin{align}S=\ft 1{16\,\pi\,G}\int\sqrt{-g}\big(R-4\,\nabla_\mu\,\Phi\nabla^\mu\Phi-\ft 1{12}\,H^2e^{-8\,\Phi}-\ft18\,k_g\,F^2\,e^{-4\,\Phi}+\frac {\delta c}3\,e^{4\,\Phi}\Big)d^3x\qquad ,\label{Action}\end{align}
describing gravity coupled to a dilaton field $\Phi$, an abelian (Maxwellian) 2-form  :
\begin{subequations}
\begin{align}F=dA\qquad,\qquad F^2=F_{\mu\nu}F^{\mu\nu}\end{align}
and  a Kalb-Ramond 3-form :
 \begin{align} H=dB-\ft {k_g}4\,A\wedge F\qquad,\qquad H^2=H_{\mu\nu\rho}H^{\mu\nu\rho}\qquad .\end{align}
 \end{subequations}
 \par\noindent
This action is invariant with respect to diffeomorphisms and gauge transformations of the potentials. Denoting  their infinitesimal generators by $\xi^\mu$, $\Lambda_\alpha$ and $\lambda$, the transformations of the various fields are given by : 
\begin{subequations}
\begin{align}&\delta g_{\mu\nu}={\cal L}_\xi g_{\mu\nu}\qquad,\label{metgaugeT}\\
& \delta B_{\alpha\beta}={\cal L}_\xi B_{\alpha\beta}+\partial_{[\alpha}\Lambda_{\beta]} -\frac {k_g}4 \partial_{[\alpha}\lambda\wedge A_{\beta]}\qquad,\label{KRgaugeT}\\
&\delta A_\alpha={\cal L}_\xi A_\alpha+\partial_\alpha \lambda\qquad, \label{MaxgaugeT}\\
&\delta \Phi={\cal L}_\xi \Phi\qquad,\label{PhigaugeT}\end{align}
\end{subequations}
where $\xi$ is a vector field, $\Lambda$ a one-form and $\lambda$ a zero-form.

Here after we make use of a natural length, that we denote by $L$, obtained  by rewriting the central charge as :
\begin{align}\delta c=:\frac{12}{L^2}\qquad.\end{align}
Our interest for the problem at hand rests on the special solution that we obtained some time ago \cite{Detournay:2005fz}, in the framework of an analysis driven by deformations of the $SL(2,{\mathbb R})$ {\sc wzw} model. Written\footnote{The subscript $(0,0)$ and superscript $(0)$ decorating the parameters $V$ and $m$ have been introduced in anticipation of more general solutions described in the following sections.} in term of a Boyer-Lindquist  retarded time  \cite{boyer:1967yu}, it is given by : 
\begin{subequations}
\begin{align}ds^2=\,&Q^2\Big(-4 \,\rho^2+V^{(0)}_{(0,0)} \,L\,\rho-\frac 14\,k_g\,L^2\big({m^{(0)}_{(0,0)}}\big)^2\Big)d\tau^2-2\,Q^2\,\rho\,d\tau\,d\rho+2\,L\,\omega\,\rho\,d\tau\,d\theta\nonumber\\
 &+\rho^2\,d\theta^2 \qquad , \label{BSmet}\\
\Phi=\,&-\ft 12\ln(\rho/L)-\ft12\ln\,Q\qquad ,\label{BSPhi}\\
A=\,&L^2\frac{m^{(0)}_{(0,0)}}{ \rho}\,d\tau+L\,c_\theta\,d\theta\qquad ,\qquad F=L^2\frac{m^{(0)}_{(0,0)}}{\rho^2}\,d\tau\wedge d\rho\label{BSMax}\\
B=\,&\frac{L^3}{Q^2\,\rho}\Big(\omega+\frac{k_g}4\,Q^2\,m^{(0)}_{(0,0)}\,c_\theta\Big)\,d\tau\wedge d\theta\qquad ,\qquad H=\omega\frac{L^3}{Q^2\,\rho^2}\,d\tau\wedge d\rho\wedge d\theta\qquad \label{BSKRB}.\end{align}
\end{subequations}
in which we use coordinates $\tau,\,\rho$ and $\theta$  whose meaning will be clarified by eqs (\ref{specoord}--\ref{simpcond}).
Of course, if the range of the $\theta$ coordinate is the all real line, by a coordinate rescaling and a gauge transformation, we always may assume that $Q=1$ and $c_\theta=0$. However if, as  from now we assume, $\theta$ is a cyclic coordinate varying between 0 and $2\,\pi$, both  $Q$ and $c_\theta$ acquire a non trivial meaning \cite{compere:2007ys} .\par\noindent
Geometrically the metric  (\ref{BSmet}) describes   a black hole, displaying  two  Killing horizons (an inner one $\rho=r_-$ and an outer one $\rho=r_+$) located at :
\begin{align}r_\pm &=\frac L {8 }\left( V^{(0)}_{(0,0)}  \pm  \sqrt{\big(V^{(0)}_{(0,0)}\big)^2 - 4\, k_g \,\big(m^{(0)}_{(0,0)} \big)^2-16\, \frac{{ \omega^2}}{Q^2} }\right)\qquad, \label{rprm}\end{align} 
assuming $\big( V^{(0)}\big)^2\geq (4 \,k_g\, \big(m^{(0)}\big)^2+16\,\omega^2)/Q^2$. In terms of $r_+$ and $r_-$ the previous field configurations, eqs (\ref{BSmet}--\ref{BSKRB}), read :
\begin{subequations}
 \begin{align}ds^2=\,&-Q^2\big(4(\rho-r_+)(\rho-r_-)\,d\tau+2\,\rho\,d\rho\big)d\tau+\big(\rho\,d\theta+L\,\omega\,d\tau\big)^2\qquad ,\label{metrpm}\\
F=\,&\frac{4\,L^2}{\sqrt{k_g}\, \rho^2}\,\sqrt{\frac{r_+\,r_-}{L^2}-\frac{\omega^2}{4\,Q^2}}\,d\tau\wedge d\rho\qquad,\qquad H=\omega\frac{L^3}{Q^2\,\rho^2}\,d\tau\wedge d\rho\wedge d\theta\qquad .\label{FHrpm}\end{align}
\end{subequations}
Asymptotically ({\it i.e. } for $\rho/L \rightarrow \infty$) the terms involving the metric parameters $V^{(0)}$, $m^{(0)}$ and $\omega$ disappear and the metric reduces to :
\begin{subequations}
 \begin{align}ds_{as}^2&=-4\,Q^2\, \rho^2\,d\tau^2-2\,Q^2\,\rho\,d\tau\,d\rho+\rho^2\,d\theta^2\label{dsas}\\
 &=L^2\,e^{\frac{2\,\sqrt{2}}Q\,x}\Big(-dt^2+dx^2+d\theta^2\Big)\qquad ,\\
 \mbox{where}\quad&:\quad \tau=\frac 1{2\,Q}\big(   t-x\big)\quad,\quad \rho=L\,e^{\frac{2}Q\,x}\qquad .\end{align}
 \end{subequations}
 It corresponds to the particular case presenting a naked singularity, obtained by setting $V^{(0)}=m^{(0)}=\omega=0$. An embedding of this geometry as submanifold of a 4 dimensional Minkowski space is described in appendix (\ref{GeomEmb}).\par\noindent
The curvature of this metric  decreases as $1/\rho^2$. However the asymptotic structure of the space is not the one of the usual 3-dimensional Minkowski space. Asymptotically the domain of this coordinate chart  is only geodesically complete with respect to space-like and null geodesics. Time-like geodesics always bounce at some maximal value (depending on the initial conditions) of the $\rho$ coordinate, just as they do on a Rindler coordinate patch. Also, in general, the angular  variable $\theta$ along the space-like and null geodesics going to infinity performs an infinite number of turns with $\rho\mapsto\infty$ : $\theta\propto\pm \ln(\rho/L)$, $\tau\propto -\ln(\rho/L)$. 
\subsection*{Field equations}
Field equations that determine the stationary configurations of the action eq. (\ref{Action}) are~:
 \begin{align}
&16\,\pi\,G\,\frac{\delta \underaccent{\dot}{L}}{\delta g_{\mu\nu}}=\sqrt{-g}\left(T^{\mu\nu}-G^{\mu\nu}\right)=:{\mathcal E}^{\mu\nu}=  0\qquad,\label{EqG} 
\end{align}
 where $T^{\mu\nu}=T_\Phi^{\mu\nu}+T_H^{\mu\nu}+T_F^{\mu\nu}+T_c^{\mu\nu}$ with :
\begin{subequations}
\begin{align}
&T_\Phi^{\mu\nu}=4\, \left(\nabla^\mu\Phi\nabla^\nu\Phi-\frac 12\, g^{\mu\nu}\,\nabla^\alpha\Phi\nabla_\alpha \Phi\right)\qquad,\\
&T_H^{\mu\nu}=\frac 14\left(H^{\mu\alpha\beta}H^\nu_{\phantom{\nu}\alpha\beta}-\frac 16\, g^{\mu\nu}\,H^{\alpha\beta\gamma}H_{\alpha\beta\gamma}\right)e^{-8\,\Phi}\qquad,\\
&T_F^{\mu\nu}=\frac {k_g}4\left(F^{\mu\alpha}F^\nu_{\phantom{\nu}\alpha}-\frac 14\, g^{\mu\nu}\,F^{\alpha\beta}F _{\alpha\beta}\right)e^{-4\,\Phi}\qquad,\\
&T_c^{\mu\nu}=\frac 16\, g^{\mu\nu}\,\delta c\,e^{4\,\Phi}\qquad,\end{align}

\end{subequations}
and
\begin{align}
 &\label{EqPhi}
16\,\pi\,G\,\frac{\delta \underaccent{\dot}{L}}{\delta \Phi}=\sqrt{-g}\left(\frac 23\,H^2\,e^{-8\,\Phi}+\frac{k_g}2\, F^2\,e^{-4\,\Phi}+\frac 43\,\delta c\,e^{4\,\Phi}+8\,\square\Phi\right)=:{\mathcal P}=  0\qquad,\\
&\label{EqA}
16\,\pi\,G\,\frac{\delta \underaccent{\dot}{L}}{\delta A_\nu}=\frac{k_g\,\sqrt{-g}}2\left(\nabla_\mu\big(e^{-4\,\Phi} \,F^{\mu\nu} -\frac 12\,e^{-{8\,\Phi}} \,H^{\mu\nu\rho}A_\rho\big)+\frac 14\,e^{-{8\,\Phi}}\, H^{\nu\mu\rho}F_{\mu\rho}\right)=:\mathcal J^\nu=  0\qquad,\\
 &\label{EqB}
16\,\pi\,G\,\frac{\delta \underaccent{\dot}{L}}{\delta {B_{\mu\nu}}}=\frac{\sqrt{-g}}2\,\nabla_\alpha\left(e^{-8\,\Phi} \,H^{\alpha\mu\nu}\right)=:{\cal K}^{\mu\nu}=  0\qquad.
\end{align} 

Before proceeding further note that :
\begin{itemize}
\item From the last equation above, we see that the abelian field equations eq.(\ref{EqA}) may be replaced by
\begin{align}\mathcal J^\nu=\frac{k_g\,\sqrt{-g}}2\,\nabla_\mu\big(e^{-4\,\Phi} \,F^{\mu\nu} -e^{-{8\,\Phi}} \,H^{\mu\nu\rho}A_\rho\big) =  0\qquad,\label{EqAbis}
\end{align}\\
\item As the space-time dimension is 3, the Kalb-Ramond field equation  (\ref{EqB}) is trivially solved in terms of a constant $\omega$, the dilaton field and the volume form $\eta=\sqrt{-g}\,dx^0\wedge dx^1\wedge dx^2$ :
\begin{align}H_{\mu\,\nu\,\rho}=\omega \frac{e^{8\,\Phi}}{L}\,\eta_{\mu\,\nu\,\rho}\label{expH}\qquad .\end{align}
\end{itemize}
To solve the remaining equations we start by expressing the fields in Bondi gauge, i.e.  
 by choosing coordinates (denoted now by $u,\, r $ and $\phi$) obtained by imposing the following gauge conditions :
\begin{align}(g_{\mu\nu})= 
\left(
\begin{array}{ccc}
g_{uu}  & g_{u\,r}  &g_{u\phi}   \\
  g_{u\,r}&  0 & 0  \\
 g_{u\phi} &0   &  r^2 
\end{array}
\right)\qquad,\qquad 
A_r=0\qquad,\qquad B_{u\,r}=B_{\phi\,r}=0\qquad .\label{gaugecond}
\end{align}
To perform the integration of the field equations we
follow {\it mutatis mutandis} the road exposed in ref.\cite{Barnich:2015jua}. We assume $u$ and $\phi$ to be dimensionless time and angular coordinates  and $r$ a radial coordinate  (of dimension $L$) and adopt their parametrisation of the metric~:
\begin{align}&g_{uu}:=r^2\,U^2(u,r,\phi)+L^2\,e^{\beta(u,r,\phi)}\,V(u,r,\phi) \ ,\  g_{u\,r}:=-L \,e^{\beta(u,r,\phi)}  \ ,\  g_{u\phi}:=r^2\,U(u,r,\phi) \ ,\nonumber \\
&g^{ur}:=-e^{-\beta(u,r,\phi)}/L\ ,\  g^{rr}:=-e^{-\beta(u,r,\phi)} \,V(u,r,\phi) \ ,\  g^{r\phi}:=e^{-\beta(u,r,\phi)} \,U(u,r,\phi)/L \label{defUVbeta}\ .\end{align}
Moreover, we rewrite the dilaton as :
\begin{align}\Phi=:-\ft12\ln(r/L)+f(u,r,\phi)\label{fdef}\qquad .
\end{align}
 
We proceed to integrate the field equations in two steps. First we show that all the  metric components and the temporal component of the Maxwell field $(A_u)$ can be expressed in terms of radial integrals involving only the dilatonic field $f(u,r,\phi)$, the abelian potential component $A_\phi(u,r,\phi)$ and five, {\it a priori}, non-trivial functions of the $u$ and $\phi$. At this stage the field equations ${\mathcal E}_{r\,r}= 0$, $\mathcal J^u= 0$, ${\mathcal E}_{r\,\phi}= 0$, ${\mathcal E}_{r\,u}= 0$ are satisfied. 
If moreover the dilaton and the $\phi$ component Maxwell equation are verified (${\mathcal P}=0$ and $\mathcal J^\phi=0$), thanks to the Bianchi identities we only have to require that the remaining equations $\mathcal J^r=0$, ${\mathcal E}_{u\,\phi}=0$, and ${\mathcal E}_{u\,u}=0$ are satisfied for a single value of the radial coordinate, while the remaining equation ${\mathcal E}_{\phi\,\phi}=0$ is then automatically satisfied.

Thus we shall first obtain the expressions of the various functions  in terms of the dilaton and the relevant component of the Maxwell field. Then we introduce an asymptotic expansion scheme and start to integrate the hierarchy  of equations it generates. In the very first steps of the procedure we shall see that the remaining arbitrariness that the Bondi gauge conditions left are just what will allows to drastically simplify the remaining equations. Going back and forth between the equations, we shall obtain an infinite triangular system of inhomogeneous  equations, involving a ``massive heat differential operator'' on the circle. At this stage the complete integration is reduced to a (cumbersome but elementary) algebraic problem. It is the occurence of this heat operator that will made the evolution of the generic solution blowing up exponentially in time (both in the past and in the future), rendering the black string solution unstable.

Of course, the difficult task of proving the convergence of the all procedure will not even be evoked in what follows (see for instance ref. \cite{Kroon:2000} for a discussion of the convergence of polyhomogeneous expansions of zero-rest-mass fields in asymptotically flat spacetimes). However a truncate closed subset of the solutions so constructed  will be exhibited, showing that the procedure could made sense.

\subsection*{Elementary integrations solving : ${\mathcal E}_{r\,r}= 0$, $\mathcal J^u= 0$, ${\mathcal E}_{r\,\phi}= 0$, ${\mathcal E}_{r\,  u}= 0$}

All functions depend on the coordinates $u,r,\phi$ (however we didn't always written them explicitly  in the expressions that follow). Derivatives with respect to $\phi$ are denoted by a prime ($F':=\partial_\phi F$), those with respect to $u$ by a dot ($\dot F:=\partial_uF$). The $r$ integrals that follows are indefinite. Thus, arbitrary functions of $u$ and $\phi$ appear as integration constants; they will be written  explicitly.
\begin{itemize}
\item From equation ${\mathcal E}_{r\,r}=  0$ we obtain :
\begin{align}\beta=\,&\int \Big(\frac{k_g}{4\,L^2}\,r\,  e^{-4\, f}\,\big(\partial_r A_\phi\big)^2+\frac 1{r}\big(1-2\,r\,\partial_r f\big)^2\Big)\,dr+\beta_{(0,0)}(u,\phi)\label{expbeta}\end{align}
\item From $\mathcal J^{u}=  0$, and setting ${\mathcal F}^{\mu\nu}:=\sqrt{g}\big(e^{-4\,\Phi}\,F^{\mu\nu}-\frac {\omega}{L^2}\eta^{\mu\,\nu\,\rho}A_\rho\big)$ we obtain :
\begin{subequations}
\begin{align}m:=\,&{\mathcal F}^{r\,u}=e^{-(4\,f+\beta)} \frac {r^3} {L^3}\big(U\,\partial_rA_\phi-\partial_rA_u\big)-\frac{\omega}{L }A_\phi\label{cFru}\qquad ,\\
=\,&-\int \frac r{L^2}\,\big(e^{-4\,f}\,\partial_rA_\phi \big)^\prime dr+m_{(0,0)}(u,\phi)\label{expm}\end{align}\qquad .\end{subequations}
\begin{align}
A_u=\,&\int \Big(U\,\partial_r\,A_\phi-\frac {L^3}{r^3}\,e^{4\,f+\beta}\,( m+\frac \omega L\,A_\phi)\Big)dr+L\,a_{u(0,0)}(u,\phi)\label{expAu}\qquad .\end{align}
\item From ${\mathcal E}_{r\,\phi}=  0$, we obtain :
\begin{subequations}
\begin{align}n:=\,&r^2\,e^{-\beta}\partial_rU\\
=\,&\frac 1 r\int\Big(4\,L\,\big(1-2\,r\,\partial_rf)f'+L\big(\beta'-r\,\partial_r\beta'\big)
+\frac {k_g}2\big(L\, m+\omega\,A_\phi\big)\partial_rA_\phi\Big)dr{\nonumber  }\\
&+\frac {L^2}r\,n_{(0,1)}(u,\phi)\label{expn}\qquad ,
\end{align}
\end{subequations}
\begin{align}
U=\,&\int \frac{e^\beta}{r^2}\,n\,dr+U_{(0,0)}(u,\phi)\label{expu}\qquad .\end{align}
\item Finally from ${\mathcal E}_{u\,r}=  0$, and using $g^{u\,r}=-e^{\beta}/L$, (along with the fact that $g_{r\,r}=0$ and $g_{r\,\phi}=0$ implies,  using the gravitational field equations, that $R_{r\,r}=T_{r\,r} $ and $R_{r\,\phi}=T_{r\,\phi} $) we obtain :
 \begin{align}V=\,&\int \Big(e^{8\,f+\beta}\,\frac {\omega^2\,L^2}{2\,r^3}+e^{4\,f+\beta}\big[-\frac 4 r+\frac {k_g}{4\,r^3}(L\,m+\omega\,A_\phi)^2\big]+\frac{e^{\beta}}r\big[4\,(f')^2+\frac 12(\beta')^2+\beta''\big]{\nonumber  }\\
&+\frac{ e^{\beta}}{2\,L^2\,r}\,n^2+\frac 1L\big(2\,U+\frac 1 r\,e^{\beta}\,n\big)'\Big)dr+V_{(0,0)}(u,\phi)\label{expV}\end{align}
\item The calculation of the asymptotic charges requires the knowledge of the Kalb-Ramond field. In the Bondi gauge that we adopt, its only non-zero component is given by :
\begin{align}B_{\phi\,u}=\,&\int \Big(e^{8\,f+\beta}\,\frac{\omega\,L^4}{r^3}+\ft 14k_g\big(A_u\,F_{r\,\phi}+A_\phi\,F_{u\,r}\big)\Big)dr+L^2\,b_{\phi\,u(0,0)}(u,\phi)\qquad ,\label{expB}\end{align}
\end{itemize}
 
\subsection*{Global electric and Kalb-Ramond charges and asymptotic expansions}

The first assumption that we will impose on the asymptotic behaviour of the fields is the requirement that the black string solution eqs (\ref{BSmet}--\ref{BSKRB}) belongs to the phase space we are looking for. Accordingly we assume as asymptotic fall-off conditions on the geometry :
\begin{align}&
 U(u,r,\phi)={\mathcal O}(1)\quad,\quad V(u,r,\phi)={\mathcal O}(r)\quad,\quad \beta(u,r,\phi)=\ln(r/L)+{\mathcal O}(1)\quad .\label{ordres}\end{align}
 and on the dilaton field :
 \begin{align}f(u,r,\phi)={\mathcal O}(1)\quad.\label{ordref}\end{align}
 As a consequence, we obtain that asymptotically  $\sqrt{-g}={\mathcal O}(r^2)$. 
 
 From eq. (\ref{EqAbis}), we infer that, on shell, an electric charge may be defined as the integral on a circle\footnote{More generally, on any closed curve of winding number 1.} $u=u_\star, \ r=r_\star$ :
 \begin{align}{\mathcal Q}_{M}=\frac 1{16\,\pi\,G}\oint_{C[u_\star,\,r_\star]}\frac 12\,k_g\,{\mathcal F}^{\alpha\beta}\epsilon_{\alpha\,\beta\,\gamma}dx^\gamma= \frac 1{16\,\pi\,G}\int_0^{2\,\pi}k_g\,{\mathcal F}^{u\,r}[u_\star,r_\star,\phi]\,d\phi\qquad .\end{align}
Equation (\ref{expm}) shows that this indeed make sense (is independent of $r_\star$); we obtain~:
\begin{align} {\mathcal Q}_{M}=-\frac 1{16\,\pi\,G}\int_0^{2\,\pi}k_g\,m_{(0,0)}(u_\star,\phi)\,d\phi\label{QMax} \end{align}
while Stokes theorem  implies that the coefficient of the zero mode in the Fourier expansion of $-k_g/(8\,G)\,m_{(0,0)}(u,\phi)$ must be a constant, see eq. (\ref{solm00}), equal to the so defined conserved electric charge.\par\noindent
Moreover a (non {\it a priori} conserved) ``topological charge'' (specific to the 3-dimensional context we are considering) may also be defined from the Maxwell field : 
\begin{align}{\mathcal Q}_{W}(u_\star)=\frac 1{16\,\pi\,G}\int_{C [u_\star,\,\infty]}A_\phi\,d\phi\label{Qhol}\end{align}
where $\int_{C[u_\star,\,\infty]}$ denotes the integral on a circle at infinity, the limit for $r_\star\rightarrow \infty$ of the same integral evaluated on the circles $C [u_\star,\,r_\star]$. As this only make sense if this limit exists, we assume henceforth that there exists a gauge such that the asymptotic behaviour of the $\phi$ component of the abelian potential reduces to :
\begin{align}A_\phi={\mathcal O}(1)\label{asympAphi}\qquad .\end{align}\par\noindent
Finally, the constant $\omega$ in eq. (\ref{expH}) may also be seen as a conserved charge. It is obtained by  integrating eq. (\ref{EqB}) contracted with the component of a closed, but non exact 1-form,  i.e. with
\begin{align}\Lambda=\Lambda_\mu\,dx^\mu=dF(u,r,\phi)+\Lambda_\phi^{(0)}\,d\phi \qquad,\qquad \Lambda_\phi^{(0)}=Cte\qquad ,\label{reducLambda}\end{align} which leads to 
 \begin{align}\nabla_\alpha\big(e^{-8\,\Phi}\,H^{\alpha\,\mu\,\beta }\Lambda_\beta\big)=0\qquad,\end{align}
 and to a Kalb-Ramond 3-D charge, defined similarly as the electric charge (see ref. \cite{compere:2007ys} for a general discussion involving $p$-forms in arbitrary dimensions), by integration on a circle :
 \begin{align}\Lambda_\phi^{(0)}\,{\mathcal Q}_{KR}&:=\frac 1{16\,\pi\,G}\oint_{C[u_\star,\,r_\star]}{\frac 12e^{-8\,\Phi}}\,H^{\alpha\, \beta\,\mu }\Lambda_\mu\,\eta_{\alpha\beta\gamma}dx^\gamma\nonumber\\
 &=\frac 1{16\,\pi\,G}\int_0^{2\,\pi}\sqrt{-g}\,e^{-8\,\Phi}\,H^{u\,r \,\phi}\Lambda_\phi^{(0)}\, dx^\phi=-\frac \omega {8\,G\,L}\,\Lambda_\phi^{(0)} \quad .\label{QKR}\end{align}
 
 All these considerations lead us to adopt the following asymptotic behaviour for the two main functions that drive the all dynamic :
\begin{align}f(u,r,\phi)=\,&\sum_{p=0}^NL^p\sum_{q=0}^p f_{(q,p)}(u,\phi)\frac {\ln^q[r/L] }{r^p}+{\mathcal O}\Big(L^{N+1}\frac{\ln^{N+1}[r/L] }{r^{N+1}}\Big)\label{asf}\qquad ,\\
A_\phi(u,r,\phi)=\,&L\,\sum_{p=0}^NL^p\sum_{q=0}^p a_{\phi(q,p)}(u,\phi)\frac {\ln^q[r/L] }{r^p}+{\mathcal O}\Big(L^{N+1}\frac{\ln^{N+1}[r/L]}{r^{N+1}}\Big)\label{asAphi}\qquad .\end{align}

\subsection*{Two fundamental simplifying relations}
Inserting these   ansatz into eqs (\ref{expbeta}--\ref{expu}) we obtain as leading terms of the asymptotic expansions of $\beta$, $A_u$ and $U$ :
\begin{align}\beta=\,&\ln(r/L)+\beta_{(0,0)}-4\frac Lr\big(\ln(r/L)f_{(1,1)}+f_{(0,1)}\big)+{\mathcal O}\Big(L^2\,\ln^2(r/L)/r^2\Big)\qquad ,\\
A_u=\,&L\,a_{u(0,0)}+\frac {L^2}r\Big( \ft12\,\ln^2(r/L)\big(\partial_\phi (e^{\beta{(0,0)}}\,a_{\phi(1,1)} )+e^{\beta{(0,0)}}\,a_{\phi(1,1)} \partial_\phi(4\,f_{(0,0)}+\beta_{(0,0)})\big){\nonumber  }\\
&  +\ln(r/L)\big(U_{(0,0)}\,a_{\phi(1,1)}+\partial_\phi (e^{\beta{(0,0)}}\,a_{\phi(1,1)} )+e^{\beta{(0,0)}}\,a_{\phi(1,1)} \partial_\phi(4\,f_{(0,0)}+\beta_{(0,0)})\big){\nonumber  }\\
&+ U_{(0,0)}\,a_{(0,1)}+ e^{4\,f_{(0,0)}+\beta_{(0,0)}
}(m_{(0,0)}+\omega\,a_{(0,0)}){\nonumber  }\\
&+\partial_\phi (e^{\beta{(0,0)}}\,a_{\phi(1,1)} )+e^{\beta{(0,0)}}\,a_{\phi(1,1)} \partial_\phi(4\,f_{(0,0)}+\beta_{(0,0)})\big)\Big) +{\mathcal O}\Big(L^3\,\ln^3(r/L)/r^2\Big)\qquad ,\label{Auexpansion}\\
U=\,&e^{\beta_{(0,0)}}\big(4\,f'_{(0,0)} +\beta'_{(0,0)}\big)\,\ln(r/L)+U_{(0,0)} +{\mathcal O}\Big(L \,\ln^2(r/L)/r\Big) \qquad .
\end{align}

Accordingly, in order to satisfy the first of the asymptotic conditions eq. (\ref{ordres}) we have to impose that :
\begin{align} \beta_{(0,0)}(u,\phi)=-4\,f_{(0,0)}(u,\,\phi)+b_{(0,0)}(u)\qquad .\label{rmkeq}\end{align}

On the other hand, solving the dilaton equation : ${\mathcal P}=0$, eq. (\ref{EqPhi}), at order $1$ (the leading order for this equation) we obtain the condition :
\begin{align}U_{(0,0)}^\prime(u,\phi)=\,& -2\,U_{(0,0)}(u,\phi)\,{f}_{(0,0)}^\prime(u,\phi)+2\,\dot {f}_{(0,0)}(u,\phi)\nonumber \\
&+ e^{b_{(0,0)}(u)-4\,{f}_{(0,0)}(u,\phi)}\big(2\,{f}_{(0,0)}^{\prime\prime}(u,\phi)-4\,({f}_{(0,0)}^\prime(u,\phi)\big)^2\big)\label{EqUgen}\qquad .\end{align}

The general solution of this equation reads  :
\begin{align} &U_{(0,0)}(u,\phi)=2\, e^{b_{(0,0)}(u)-4\,{f}_{(0,0)}(u,\phi)}\,{f}_{(0,0)}^\prime(u,\phi)  +e^{-2\,{f}_{(0,0)}(u,\phi)}\partial_uF(u,\phi) \label{Ugen}
 \\
 &  \mbox{with}\qquad F(u,\phi)=\int_0^\phi e^{2\,{f}_{(0,0)}(u,\xi)}d\xi +h(u)\label{GenF}\qquad .\end{align}
 
But in order for  $U_{(0,0)}(u,\phi)$ to be periodic, i.e. to have $ U_{(0,0)}(u,0)= U_{(0,0)}(u,2\,\pi)$, we  require that :
\begin{align}\partial_u \int_0^{2\,\pi} e^{2\,{f}_{(0,0)}(u,\phi)}d\phi=0\qquad ,\label{dQ0}\end{align}
or equivalently that :
\begin{align}\int_0^{2\,\pi} e^{2\,{f}_{(0,0)}(u,\xi)}d\xi=2\,\pi/Q\qquad ,\end{align}
where $Q$ is a (positive) constant.

Here comes a crucial simplification for the continuation of our discussion on the asymptotic solutions. Since the Bondi gauge conditions do not completely fix the coordinates, we may still perform two kinds of diffeomorphisms :
\begin{align} \tau=T(u)\qquad\mbox{and}\qquad \rho=r/\partial_\phi H(u,\,\phi)\quad,\quad \theta =H(u,\,\phi)\qquad \label{diffeoTH}.\end{align}
Of course to be well defined we have to impose the usual conditions (assuming a choice of the orientations of the time and angular coordinates)~:
\begin{align}\partial_uT(u)>0\qquad,\qquad\partial_\phi H(u,\,\phi)>0\qquad,\qquad H(u,\,2\,\pi)-H(u,\,0)=2\,\pi\qquad ;\end{align}
 we remind that both $\phi$ and $\theta$ are assumed to be cyclic variables of period $2\,\pi$.\par\noindent
These conditions are exactly fulfilled by choosing $T(u)$ and $H(u,\,\phi)$ as :
\begin{align}T(u)=\int_0^u e^{b_0(\upsilon)}\,d\upsilon\qquad\mbox{and}\qquad H(u,\,\phi)=Q\,F(u,\,\phi)\qquad .\label{TebHQF}\end{align}
Thus by redefining the coordinates as :
\begin{align}\tau=\int_0^u e^{b_{(0,0)}(\upsilon)}\,d\upsilon\qquad,\qquad\rho = e^{-2\,{f}_{(0,0)}(u,\phi)}\,r/Q\qquad,\qquad \theta= Q\,F(u,\phi)\qquad,\label{specoord}\end{align}
we see that we may always assume that we are working in a coordinate system $\{\tau,\,\rho,\,\theta\}$ such that :
\begin{align}{f}_{(0,0)}(\tau,\theta)= -\ft 12 \ln(Q)\  ,\  b_{(0,0)}(\tau)=0\  i.e.\  \beta_{(0,0)}(\tau,\theta)=2\, \ln(Q)\ ,\ 
U_{(0,0)}(\tau,\theta)=0\ .\label{simpcond} \end{align}
Here $Q$ is a constant   and the new angular coordinate, denoted by $\theta$, still varys between $0$ and $2\,\pi$. It is on such a chart that actually the black string geometry  is  described by eqs (\ref{BSmet}--\ref{BSKRB}). A glance on the formulae that govern the transformation of the asymptotic expansions of the various unknown functions   that have to be evaluated convinces rapidly on the usefulness to work on a $\{\tau,\,\rho,\,\theta\}$ chart, where
eqs (\ref{simpcond}) are satisfyed.
 We could even assume that $Q=1$ and restoring it at the end of the integration by a simple rescaling of the cyclic variable and an appropriate redefinition on the summation index in the Fourier expansion of the various functions.
 
 Note that  all these diffeomorphisms  are usually considered as being large diffeomorphisms. Indeed  the expressions of the asymptotic Killing vectors we shall consider later, when discussing the asymptotic charges, are not left invariant under their action. Thus, in principle, the values of the various charges they define change when such diffeomorphisms act on a physical configuration : large diffeomorphisms transform  into distinct dynamical states of the physical system. But this conclusion presuppose that after a large diffeomorphism charges associated to the reducibility parameters given by their original expressions are still defined. For the problem we consider here it will be shown that this is not the case.

\section{Integration of the asymptotic expansions\label{IntasympEqs}}

In this section we  describe the path we have followed to integrate order by order the remaining field equations. We almost completely fix the gauge by imposing the choices   (\ref{simpcond}). The remaining freedom only consists into the choice of the lower bound of the integrals defining $T(u)$ and $F(u,\phi)$, corresponding to translations of the $\tau$ and $\theta$ variables. \\
  The equations we have to consider consist of two standard evolution equations $\mathcal J^\theta =0$ and ${\mathcal P}=0$ governing the evolution of $f$ and $A_\theta$ and three complementary equations that will fix the integration functions $m_{(0,0)}(u,\,\phi)$, $n_{(0,1)}(u,\,\phi)$ and $V_{(0,0)}(u,\,\phi)$, introduced in eqs (\ref{expm}, \ref{expn}, \ref{expV}). \\
We start with the expressions of ${\beta}_{(0,0)}({\tau},\,\theta)=-2\,\ln(Q)$ and  $  U_{(0,0)}({\tau},\,\theta)=0$ obtained in eqs (\ref{rmkeq}, \ref{simpcond}), that solve the equation ${\mathcal P} =0$ at order one. From now we  shall proceed  order by order, taking into account at each order the results established at the previous ones and represent $\tau$-derivatives by   dots and $\theta$-derivatives by   primes.
\vspace{2mm}
\par  At  order ($1$), the equation ${\mathcal J}^\theta=0$ implies that   :
\begin{align}{a}_{{\tau}(0,0)}({\tau},\,\theta)=\dot{{\lambda}}({\tau},\,\theta)\qquad,\qquad {a}_{\theta(0,0)}({\tau},\,\theta)= {\lambda}'({\tau},\,\theta)+c_\theta\qquad, \label{A00}\end{align}
  where $\ c_\theta$ is a constant.
In other words these components reflect the expected residual gauge freedom that preserves the Bondi gauge but also encode the topological charge eq.(\ref{Qhol}) that, as a consequence of eq. (\ref{asympAphi}), defines  a conserved charge.\par
At order $\ln(\rho/L)/{\rho}$ and $1/{\rho}$, respectively, we obtain from the dilaton equation $\mathcal P=0$ (\ref{EqPhi}):
\begin{subequations}
\begin{align}\dot {f}_{(1,1)}({\tau},\,\theta)=\,& -Q^2\,{f}_{(1,1)}^{\prime\prime}({\tau},\,\theta)\qquad,\label{Jacf11}\\
\dot {f}_{(0,1)}({\tau},\,\theta)=\,& -Q^2\,{f}_{(0,1)}^{\prime\prime}({\tau},\,\theta) 
+2\,\dot {f}_{(1,1)}({\tau},\,\theta)+4\,{f}_{(1,1)}({\tau},\,\theta)\qquad,\label{Jacf01}\end{align} 
and from the Maxwell equation $\mathcal J^\theta = 0$ (\ref{EqA}):
\begin{align}\dot {a}_{\theta(1,1)}({\tau},\,\theta)=\,&-Q^2\,{a}_{\theta(1,1)}^{\prime\prime}({\tau},\,\theta)\qquad ,\label{Jaca11}\\
\dot {a}_{\theta(0,1)}({\tau},\,\theta)=\,&-Q^2\,{a}_{\theta(0,1)}^{\prime\prime}({\tau},\,\theta)+2\,\dot {a}_{\theta(1,1)}({\tau},\,\theta)+4\,{a}_{\theta(1,1)}({\tau},\,\theta) \label{Jaca01}\qquad .\end{align} 
\end{subequations}
\par   At order ${\rho}^{-2}$ (in the field equations), we obtain nine equations :  
\begin{itemize}
\begin{subequations}
\item {From $\mathcal P=0$ at order $ \ln^4({\rho}/L)/{\rho}^{2} $} :
\begin{align}3\,\big({f}_{(1,1)}^\prime({\tau},\,\theta)\big)^2-2\,{f}_{(1,1)}({\tau},\,\theta)\,{f}^{\prime\prime}_{(1,1)}({\tau},\,\theta)=0\qquad,\end{align}
whose only periodic solution compatible with eq. (\ref{Jacf11}) is
\begin{align}{f}^\prime_{(1,1)}({\tau},\,\theta)=0\quad \mbox{i.e.}\quad {f}_{(1,1)}({\tau},\,\theta)={f}^{(0)}_{(1,1)}\label{solf11}\qquad,\end{align}
 with ${f}^{(0)}_{(1,1)}$ an arbitrary constant.
\item As a direct consequence of eq. (\ref{solf11}), equation $\mathcal J^\theta=0$ :
\begin{align}3\,a_{\theta(1,1)}'(\tau,\,\theta)\, f'_{(1,1)}(\tau,\,\theta)-2 \,a_{\theta(1,1)}(\tau,\,\theta)\, f"_{(1,1)}(\tau,\,\theta)=0\end{align}
 is satisfied at order $ \ln^4({\rho}/L)/{\rho}^{2} $.
 \end{subequations}
\item {From $\mathcal J^\theta=0$ at order $ \ln^3({\rho}/L)/{\rho^2} $} :
\begin{subequations}
\begin{align}{a}_{\theta(1,1)} ({\tau},\,\theta)\,{f}^{\prime\prime}_{(0,1)}({\tau},\,\theta)-{a}_{\theta(1,1)}^\prime({\tau},\,\theta)\,{f}^\prime_{(0,1)}({\tau},\,\theta)={f}^{(0)}_{(1,1)} \,{a}_{\theta(1,1)} ^{\prime\prime}({\tau},\,\theta)\label{caf11}\qquad .\end{align}
\item {From $\mathcal P=0$ at order $ \ln^3({\rho}/L)/{\rho}^{2} $ :}
\begin{align}\big({a}_{\theta(1,1)}^\prime({\tau},\,\theta)\big)^2-{a}_{\theta(1,1)} ({\tau},\,\theta)\,{a}^{\prime\prime}_{\theta(1,1)}({\tau},\,\theta)=\frac {32}{k_g\,Q^2}\,{f}^{(0)}_{(1,1)}\,{f}_{(0,1)} ^{\prime\prime}({\tau},\,\theta)\label{ca11}\qquad .\end{align}
\end{subequations}
\end{itemize}
\begin{subequations}
Assuming $k_g>0$ , eqs (\ref{caf11}, \ref{ca11}) imply that\footnote{Set $\kappa=\sqrt{2\,k_g}\,Q/8$, $k=j_{(1,1)}\,\kappa$, $f_{(0,1)}=\kappa \,f$, $\zeta=a_{\theta(1,1)}+i\,f$, $\xi=(k+i\,a_{\theta(1,1)})$. Equations (\ref{caf11}, \ref{ca11}) imply that $\zeta'\,\xi'=\xi\,\zeta''$, from which we conclude. Note that if $k_g<0$ other branches are possible, but then ghosts appear in the action [eq.(\ref{Action})].}:
\begin{align}{a}^\prime_{(1,1)}({\tau},\,\theta)=0\quad\mbox{i.e. }\quad{a}_{\theta(1,1)}({\tau},\,\theta)={a}^{(0)}_{\theta(1,1)} \quad  \label{sola11}\end{align}
with ${a}^{(0)}_{\theta(1,1)}$ a constant, see eq. (\ref{Jaca11}),  and :
\begin{align}{f}_{(0,1)}({\tau},\,\theta)={j}_{(0,1)}(\tau)\qquad \mbox{or}\qquad {f}^{(0)}_{(1,1)}=0\qquad.\end{align}
\end{subequations}
To proceed further let us restrict ourselves to the dynamical solution sector and   choose the solution ${f}^{(0)}_{(1,1)}=0$.\\
Then, the  solutions of eqs (\ref{Jaca01},\ref{Jacf01}) are given by the superpositions of  modes~:
\begin{subequations}
 \begin{align}{f}_{ (0,1)}({\tau},\,\theta)=\,&\sum_n {f}_{ (0,1)}^{(n)}e^{i\,n{\,\theta\,}+Q^2\,n^2\,\tau} \label{solf01}\qquad ,\\
 {a}_{\theta(0,1)}({\tau},\,\theta)=\,&\sum_n {a}_{\theta(0,1)}^{(n)}e^{i\,n{\,\theta\,}+Q^2\,n^2\,\tau}+4\,\tau\,{a}^{(0)}_{\theta(1,1)} \label{sola01}\qquad .\end{align}
 \end{subequations}
Of course we have to impose the reality conditions (always understood in the following and that we shall not repeat explicitly) :
\begin{equation}
{f}_{ (0,1)}^{(n)}=\big({f}_{ (0,1)}^{(-n)}\big)^\star\qquad,\qquad {a}_{\theta(0,1)}^{(n)}=\big({a}_{\theta(0,1)}^{(-n)}\big)^\star\qquad.\end{equation}

 These modes are obviously unstable : they blow up exponentially when $\tau$ increases.\par
 But this is not the end of the story. It is interesting to pursue the discussion to next order.
  \begin{itemize}
  \begin{subequations}
 \item {From the equation $\mathcal P=0$, at order $ \ln^2({\rho}/L)/{\rho}^{2} $}, we obtain :
 \begin{align} \dot {f}_{(2,2)}({\tau},\,\theta)=&  -\frac{Q^2}3\,{f}_{(2,2)}^{\prime\prime}({\tau},\,\theta)-\frac 83\,{f}_{(2,2)}({\tau},\,\theta)+\frac 23 Q^2\Big({f}_{(0,1)}({\tau},\,\theta){f}_{(0,1)}^{\prime\prime}({\tau},\,\theta)-2\big({f}_{(0,1)}^{\prime }({\tau},\,\theta)\big)^2\Big) {\nonumber  }\\
&
 +\frac 1{24}\,k_g\,Q^2\Big(Q^2\,{a}^{(0)}_{(1,1)}\,{a}_{\theta(0,1)}^{\prime\prime}({\tau},\,\theta)-4\,\big( {a}^{(0)}_{(1,1)}\big)^2\Big)\qquad  .\label{Jacf22}\end{align} 
  \item {From the equation $\mathcal J^\theta=0$, at order $ \ln^2({\rho}/L)/{\rho}^{2}$} :
  \begin{align}&\dot {a}_{\theta(2,2)}({\tau},\,\theta)=-\frac{Q^2}3\,{a}_{\theta(2,2)}^{\prime\prime}({\tau},\,\theta)-\frac 83\,{a}_{\theta(2,2)}({\tau},\,\theta){\nonumber  }\\
&+\frac 23 Q^2\Big(({a}_{\theta(0,1)}({\tau},\,\theta)-{a}^{(0)}_{(1,1)}){f}_{(0,1)}^{\prime\prime}({\tau},\,\theta)-2\,{a}_{\theta(0,1)}^{ \prime}({\tau},\,\theta)\,{f}_{(0,1)}^{ \prime}({\tau},\,\theta) 
-\frac 1{2}   \,{a}^{(0)}_{(1,1)}\,{n}_{(0,1)}^{\prime }({\tau},\,\theta)\Big) \ .\label{Jaca22}
\end{align} 
\end{subequations}
\item {The equation $\mathcal P=0$, at order $ \ln({\rho}/L)/{\rho}^{2}$}, leads to a similar equation but   involving the not yet fixed functions $m_{(0,0)}(\tau,\,\theta)$ and $n_{(0,1)}(\tau,\,\theta)$ :
\begin{subequations}
\begin{align}&\dot {f}_{(1,2)}({\tau},\,\theta)=-\frac{Q^2}3\,{f}_{(1,2)}^{\prime\prime}({\tau},\,\theta)-\frac 83\,{f}_{(1,2)}({\tau},\,\theta) {\nonumber  }\\
&+Q^2\Big(\frac{k_g}{24}\,Q^2\big(({a}_{\theta(0,1)}({\tau},\,\theta)+\frac {17}{6}\,{{a}^{(0)}_{(1,1)}}){a}_{\theta(0,1)}({\tau},\,\theta)^{\prime\prime}
-({a}_{\theta(0,1)}({\tau},\,\theta)^{\prime})^2\big) 
-\frac{k_g}3\,{{a}^{(0)}_{(1,1)}}\,{a}_{\theta(0,1)}({\tau},\,\theta)\Big){\nonumber  }\\
&+\frac {4\,Q^2}9\Big(2\,{f}_{(0,1)} ({\tau},\,\theta){f}_{(0,1)}^{\prime\prime}({\tau},\,\theta)-\big({f}_{(0,1)}^\prime({\tau},\,\theta)\big)^2-{f}_{(2,2)}^{\prime\prime}({\tau},\,\theta)\Big){\nonumber  }
\\
&+\frac{Q^2}3\big(2\,{n}_{(0,1)}({\tau},\,\theta)\,{f}_{(0,1)}^{\prime }({\tau},\,\theta)-{n}_{(0,1)}^\prime({\tau},\,\theta)\,{f}_{(0,1)}({\tau},\,\theta)\big){\nonumber  }\\
&+\frac {k_g\,Q^2}{24} \, {{a}^{(0)}_{(1,1)}}\big(m_{(0,0)}^\prime({\tau},\,\theta)+\omega\, {\lambda}''({\tau},\,\theta)-\frac{16}3\,{{a}^{(0)}_{(1,1)}}\big) {\nonumber  }\\
&+\frac{40}9\,{f}_{(2,2)}({\tau},\,\theta)\label{Jacf12}\qquad .\end{align}
 
\item {Equation $\mathcal J^\theta=0$,  at the same order   $ \ln({\rho}/L)/{\rho}^{2}$,  implies that : }

\begin{align}\dot {a}_{\theta(1,2)}({\tau},\,\theta)=\,&-\frac{Q^2}3\,{a}_{\theta(1,2)}^{\prime\prime}({\tau},\,\theta)-\frac 83\,{a}_{\theta(1,2)}({\tau},\,\theta){\nonumber  }\\
&+ {Q^2} \Big( \big(\frac{14}9\,{a}_{\theta(0,1)}({\tau},\,\theta)-\frac{44}9\,{{a}^{(0)}_{(1,1)}}\big)\,{f}_{(0,1)}^{\prime\prime}({\tau},\,\theta){\nonumber  }\\
& \  -\frac 49\,{a}_{\theta(2,2)}''({\tau},\,\theta)-\frac{4}3 \,{f}_{(0,1)} ({\tau},\,\theta)\,{a}_{\theta(0,1)}''({\tau},\,\theta)+\frac 29\,f_{ (0,1)}^\prime ({\tau},\,\theta)\,{a}_{\theta(0,1)}^\prime ({\tau},\,\theta){\nonumber  }\\
& \ +\frac13\big({n}_{(0,1)} ({\tau},\,\theta)\,{a}_{\theta(0,1)}'({\tau},\,\theta)-{n}_{(0,1)}^\prime({\tau},\,\theta)({a}_{\theta(0,1)} ({\tau},\,\theta)+\frac76\,{{a}^{(0)}_{(1,1)}})\big)\Big){\nonumber  }
\end{align}
\begin{align}
&+\frac43\big(4\,{{a}^{(0)}_{(1,1)}}\,{f}_{(0,1)}({\tau},\,\theta)-(m_{(0,0)}({\tau},\,\theta)+\omega\,{\lambda}'({\tau},\,\theta)){f}_{(0,1)}'({\tau},\,\theta)\big){\nonumber  }\\
&+\frac 13\,{V}_{(0,0)}({\tau},\,\theta)\,{{a}^{(0)}_{(1,1)}}
+\frac 13\,\omega\,{a}_{\theta(0,1)}'({\tau},\,\theta)-\frac {40}9\,{a}_{\theta(2,2)}({\tau},\,\theta)\qquad .\label{Jaca12}
\end{align}

\end{subequations}

The occurence, on the righthand sides, of a term proportional to what is an arbitrary gauge parameter : $\lambda'(\tau,\,\theta)$ in eq. (\ref{Jaca12}), $\lambda''(\tau,\,\theta)$ in eq. (\ref{Jacf12})   is not surprising. Indeed, as we may see from eq. (\ref{expm}) it is the combination $m(\tau,\,\theta)+\omega\,A_\theta(\tau,\,\theta)/L$ and more specifically, in the asymptotic expansion~: $m_{(0,0)}(\tau,\,\theta)+\omega\,a_{\theta(0,0)}$ (see eq. (\ref{Auexpansion})), that is gauge invariant with respect to the residual gauge transformation that preserves the Bondi gauge. That this gauge invariant term  and its derivatives occur in equations governing subdominant terms of order  $1/\rho^2$ in the expansions of $f$ and $A_\theta$ results, as eq. (\ref{Auexpansion}) shows, from  the fact that ``$m(\tau,\,\theta)+\omega\,A_\theta(\tau,\,\theta)/L$" contributes to the order $1/\rho$ of the expansion of $A_u$: it is like an ``$a_{u(0,1)}$'' term. 
 \item {From the equation ${\mathcal P}=0$, expanded at order $ 1/{\rho}^{2}$}, we obtain :
 \begin{subequations}
\begin{align}\dot {f}_{(0,2)}({\tau},\,\theta)=\,&-\frac{Q^2}3\,{f}_{(0,2)}^{\prime\prime}
    ({\tau},\,\theta)-\frac 83\,{f}_{(0,2)}({\tau},\,\theta){\nonumber  }\\
&-Q^2\Big( \frac{2}{9}{k_g} {{a}^{(0)}_{(1,1)}}
   {a}_{\theta(0,1)}({\tau},\,\theta ){\nonumber  }\\
   &+\frac{1}{24}{k_g}
   {a}_{\theta(0,1)}'({\tau},\,\theta )\big( m_{0,0}({\tau},\,\theta )+\omega\,a_{\theta(0,0)}({\tau},\,\theta )\big){\nonumber  }\\
   &-\frac{1}{24}
  {k_g} {a}_{\theta(0,1)}({\tau},\,\theta ) \big({m}_{(0,1)}'({\tau},\,\theta 
   ) + \omega \, {a}_{\theta(0,1)}'({\tau},\,\theta )
  \big){\nonumber  }\\&
    +\frac{1}{6}{k_g} {a}_{\theta(0,1)}({\tau},\,\theta ){}^2-\frac{1}{9}
   {f}_{(0,1)}'({\tau},\,\theta) {n}_{(0,1)}({\tau},\,\theta ){\nonumber  }\\
   &-\frac{1}{9}
  {f}_{(0,1)}({\tau},\,\theta ){n}_{(0,1)}'({\tau},\,\theta )-\frac{37}{27}
   \big({f}_{(0,1)}'({\tau},\,\theta)\big)^2{\nonumber  }\\
   &-\frac{7}{27}{f}_{(0,1)}({\tau},\,\theta )
 {f}_{(0,1)}''({\tau},\,\theta )+\frac{2}{9} f_{1,2}{}^{(0,2)}({\tau},\,\theta ){\nonumber  }\\
   &+\frac{8}{27} {f}_{(0,2)}''({\tau},\,\theta )
   +\frac{5}{144}{k_g}
   {{a}^{(0)}_{(1,1)}} \big({m}_{(0,1)}'({\tau},\,\theta  )+ \omega
   \, a_{\theta(0,0)}''({\tau},\,\theta )\big){\nonumber  }\\
   &-\frac{1}{54}{k_g}
   {{a}^{(0)}_{(1,1)}}^2+\frac{1}{12} {n}_{(0,1)}^2({\tau},\,\theta )
   \Big) {\nonumber  }\\
   &-\frac{19}{216}{k_g} Q^4 {{a}^{(0)}_{(1,1)}} a_{0,1}''(\tau,\,\theta)
   -\frac{5}{72}{k_g} Q^4 \big({a}_{0,1}'({\tau},\,\theta 
   )\big)^2{\nonumber  }\\
   &+\frac{11}{72}{k_g} Q^4 {a}_{\theta(0,1)}({\tau},\,\theta )
   {a}_{0,1}''({\tau},\,\theta ){\nonumber  }\\
   &+\frac{1}{3}{f}_{(0,1)}({\tau},\,\theta )
   {V}_{(0,0)}({\tau},\,\theta )+\frac{20}{9} {f}_{(1,2)}({\tau},\,\theta
   )+\frac{8}{27} f_{(2,2)}({\tau},\,\theta)+\frac{\omega ^2}{12 \,Q^2}\qquad .\label{eqf02}\end{align}
    \item {Similarly, at order $1/{\rho}^{2}$},  the equation $\mathcal J^\theta= 0$ results in :

\begin{align}\dot {a}_{\theta(0,2)}({\tau},\,\theta)=\,&-\frac{Q^2}3\,{a}_{\theta(0,2)}^{\prime\prime}({\tau},\,\theta)-\frac 83\,{a}_{\theta(0,2)}({\tau},\,\theta){\nonumber  }\\
&- {Q^2} \Big(\frac{8}{27}
   {a}_{\theta(2,2)}''({\tau},\,\theta ) +\frac{2}{9}  {a}_{\theta(1,2)}''({\tau},\,\theta ){\nonumber  }\\
   &+\frac{52}{27} {{a}^{(0)}_{(1,1)}}
   {f}_{(0,1)}{}^{(0,2)}({\tau},\,\theta ) 
  -\frac{19}{108} {{a}^{(0)}_{(1,1)}}  
  n _{(0,1)}'({\tau},\,\theta ){\nonumber  }\\  &+\frac{26}{9} {a}_{\theta(0,1)}''({\tau},\,\theta ) {f}_{(0,1)}(\tau,\,\theta){\nonumber  }\\
   &-\frac{85}{27} {a}_{\theta(0,1)}'({\tau},\,\theta ) {f}_{(0,1)}'(\tau,\,\theta) +\frac{5}{18} {a}_{\theta(0,1)}'({\tau},\,\theta ) {n} _{(0,1)}(\tau,\,\theta)
    {\nonumber  }\\
   &+\frac{17}{27} {a}_{\theta(0,1)}({\tau},\,\theta ) {f}_{(0,1)}^{\prime\prime}(\tau,\,\theta)+\frac{1}{18} {a}_{\theta(0,1)}({\tau},\,\theta ) {n} _{(0,1)}'(\tau,\,\theta)\Big)
  {\nonumber  }\\ 
   &+\frac{8}{27} a_{2,2}({\tau},\,\theta ) +\frac{20}{9} a_{1,2}(\tau,\,\theta) +\frac{32}{9} {{a}^{(0)}_{(1,1)}}
   {f}_{(0,1)}({\tau},\,\theta ) {\nonumber  }\\
   & +\frac{7}{18} \omega\, 
   {a}_{\theta(0,1)}'({\tau},\,\theta )
   +\frac{16}{3} {a}_{\theta(0,1)}({\tau},\,\theta ) {f}_{(0,1)}({\tau},\,\theta )  
  {\nonumber  }\\&-\frac{4}{3}\big(m_{(0,0)}'({\tau},\,\theta ) + \omega\,{a_{\theta(0,0)}}''({\tau},\,\theta )\big) 
   \,  {f}_{(0,1)}({\tau},\,\theta ){\nonumber  }\\ 
      &+\big( m_{(0,0)}(u,\theta)+\omega\,a_{\theta(0,0)}'({\tau},\,\theta )\big)\big(\frac{\omega}{3 \, Q^2}+ \frac 13\, {n} _{(0,1)}(\tau,\,\theta)-\frac{8}{9}\, 
   {f}_{(0,1)}{}^{(0,1)}({\tau},\,\theta )\big)\nonumber \\ 
   &+\big(\frac{1}{3}
   {a}_{\theta(0,1)}({\tau},\,\theta )-\frac{4}{9}
   {{a}^{(0)}_{(1,1)}} \big){V}_{(0,0)}({\tau},\,\theta ) \label{eqa02} \qquad . \end{align} 
   \end{subequations}
 \end{itemize}
 \vspace{2mm}
 Before discussing the general structure of the hierarchy of equations, let us turn to the complementary equations.
 \par Solving $\mathcal J^\rho=0$.\\ 
  \begin{itemize}
  \item The order $\ln({\rho}/L)$ term of the expansion of $\mathcal J^\rho$ is proportional to :
  \begin{align}\dot {a}'_{\theta(0,1)}({\tau},\,\theta)=\,&{U}_0({\tau}) \,{a}_{\theta(0,1)}''({\tau},\,\theta)-Q^2\,{a}_{\theta(0,1)}^{\prime\prime\prime}({\tau},\,\theta)\qquad ,\end{align}
 and thus, as a consequence of eqs (\ref{Jaca01}) and   (\ref{sola11}), the corresponding equation  is satisfied at this order.
 \item At order $1$, we obtain the equation :
 \begin{align}\big(  m_{(0,0)}({\tau},\,\theta)+\omega\,{a}_{\theta(0,0)}({\tau},\,\theta)\big)\dot{\strut}=\,&-Q^2\big(m_{(0,0)}({\tau},\,\theta)+\omega\,a_{\theta(0,0)}({\tau},\,\theta)\big)''{\nonumber  }\\
 &+4\,Q^2\,a'_{\theta(0,1)}-2\,Q^4\,a^{\prime\prime\prime}_{\theta(0,1)}\qquad ,\end{align}
 whose general solution is given by a mode superposition similar to those displayed in eqs (\ref{solf01}, \ref{sola01}) :
 \begin{align}{m}_{ (0,0)}({\tau},\,\theta)+\omega\,{a}_{\theta(0,0)}({\tau},\,\theta)=\,&\sum_n {m}_{ (0,0)}^{(n)}e^{i\,n{\,\theta\,}+Q^2\,n^2\,\tau}{\nonumber  }\\&+\tau\,\big(4\,Q^2\,a'_{\theta(0,1)}-2\,Q^4\,a^{\prime\prime\prime}_{\theta(0,1)}\big)\qquad .\label{solm00}\end{align}
 So $\mathcal J^\rho$ vanishes at infinity and thus, as a consequence of the Bianchi identities, it vanishes everywhere.
\end{itemize}
\par {Solving $\rho\,{\mathcal E}_{u\phi}\simeq0$ (at one point)}.\\
Here as well, we find that the leading coefficient of the expansion of $\rho\,{\mathcal E}_{u\phi}$ (the term of order $\ln(\rho/L)$ vanishes as a consequence of eqs (\ref{Jacf01}) and (\ref{solf11}).
\begin{itemize}
\item At order 1, this equation leads to the equation :
\begin{align}\dot {n}_{(0,1)}({\tau},\,\theta)=\,& -Q^2\,{n}''_{(0,1)}({\tau},\,\theta) 
 -16 \,f'_{(0,1)}({\tau},\,\theta)+8\,Q^2\,f^{\prime\prime\prime}_{(0,1)}({\tau},\,\theta)\qquad ,\end{align}
 whose general solution reads
 \begin{align} {n}_{(0,1)}({\tau},\,\theta)=\,&\sum_p {n}_{ (0,1)}^{(p)}e^{i\,p{\,\theta\,}+Q^2\,p^2\,\tau}-\tau\,\big(16\,f'_{ (0,1)}+8\,Q^2\,f^{\prime\prime\prime}_{ (0,1)}\big)\label{soln01}\qquad .\end{align}
\end{itemize}
\vspace{2mm}
\par{The equation $\rho\,{\mathcal E}_{uu}=0$ also has to be only satisfied for a single value of $\rho$.}\\
\begin{itemize}
\item   Taking into account all previous equations, only the term of order 1 in the expansion of $\rho\,{\mathcal E}_{uu} $, provides a new condition, that fixes the last unknown function $V_{(0,0)}$. It has to satisfy the equation : 
\begin{align}\dot {V}_{(0,0)}({\tau},\,\theta)=\,&-Q^2\,{V}_{ (0,0)}''({\tau},\,\theta)
+Q^2\big(32\,{f}_{(0,1)}''({\tau},\,\theta)-4\,{n}_{(0,1)}'({\tau},\,\theta)\big){\nonumber  }\\ 
&+Q^4\big(2\,{n}_{(0,1)}^{\prime\prime\prime}({\tau},\,\theta)-24\,{f}_{(0,1)}^{\prime\prime\prime\prime}({\tau},\,\theta)\big) \qquad ,\end{align}
whose solution is :
\begin{align} {V}_{(0,0)}({\tau},\,\theta)=\,&\sum_n {V}_{ (0,0)}^{(n)}e^{i\,n\,\theta\,+Q^2\,n^2\,\tau}+\tau\,Q^2\big(32\,{f}_{(0,1)}''({\tau},\,\theta)-4\,{n}_{(0,1)}'({\tau},\,\theta)\big){\nonumber  }\\ 
&+\tau\,Q^4\big(2\,{n}_{(0,1)}^{\prime\prime\prime}({\tau},\,\theta)-24\,{f}_{(0,1)}^{\prime\prime\prime\prime}({\tau},\,\theta)\big)\label{solV00}\qquad .\end{align}
\end{itemize}
\vspace{2mm}

The general pattern for the complete integration of the field equations emerges from these first steps. It is the following : 
at each order in $1/\rho^p$ ,
the equations driving, both ${f}_{(q,p)}$ and $a_{\theta(q,p)}$ (here after generically  denoted by $X_{(q,p)}$) have the following general structure :
\begin{align}{\mathcal D}_p\,X_{(q,p)}={\mathcal S}_{(q,p)} \qquad ,\label{genafeq}\end{align}
where ${\mathcal D}_p$ is the parabolic differential operator (a massive heat flow operator) :
\begin{align}{\mathcal D}_p:=\partial_\tau+\frac {Q^2}{(2\,p-1)}\partial^2_\theta +\frac {4\,p(p-1)}{(2\,p-1)}\qquad .\end{align}
The general solution of the homogeneous equation ${\mathcal D}_p\,X_{(q,p)}=0$ is given by the mode superposition :
\begin{align} X^{hom}_{(q,p)}=\sum_nx^{(n)}_{(q,p)}e^{i\,n\,\theta+\frac{(Q^2\, n^2-4\,p\,(p-1))}{(2\,p-1)}\tau}\qquad .\label{genXqp}\end{align}
Except  for at most one of them, all the remaining modes are exponentially unstable.  Those with $n>2\, \sqrt{p(p-1)}/Q$ blow up for $\tau$ going to $+\infty$; those with $n<2\, \sqrt{p(p-1)}/Q$ diverge in the limit where $\tau\rightarrow -\infty$.\\
The right-hand side of the equation (\ref{genafeq}) is given by a sum of products of power $\tau^k$ of time variable $\tau$ and modes solving  the homogeneous equation for lower (or equal) values of the $p$ index. As product of such modes constitute in general eigenmodes of the operator ${\mathcal D}_p$ but with a non zero eigenvalue, it is immediate that the special solution of the non-homogeneous equation with the source term ${\mathcal S}_{(q,p)}$ will be given by a sum of products of the same eigenmodes multiplied by a polynomial of the  $\tau$   variable. This polynomial may be chosen to be equal to $\tau^{k+1}/(k+1)$ if the eigenmode of ${\mathcal D}_p$ is of zero eigenvalue ( for instance, see eqs (\ref{sola01}, \ref{solm00}, \ref{soln01}, \ref{solV00}) or equal to $e^{-s\,\tau}\partial_\sigma^k\big(e^{\sigma\,\tau}/\sigma\big)\vert_{\sigma=s}$, if the eigenmode is of non-zero eigenvalue $s$.  But actually this procedure is formal as it requires in general to manage products of series.
\subsection*{A special solution}
\begin{figure}[ht]
\begin{center}
\includegraphics[height=40mm]{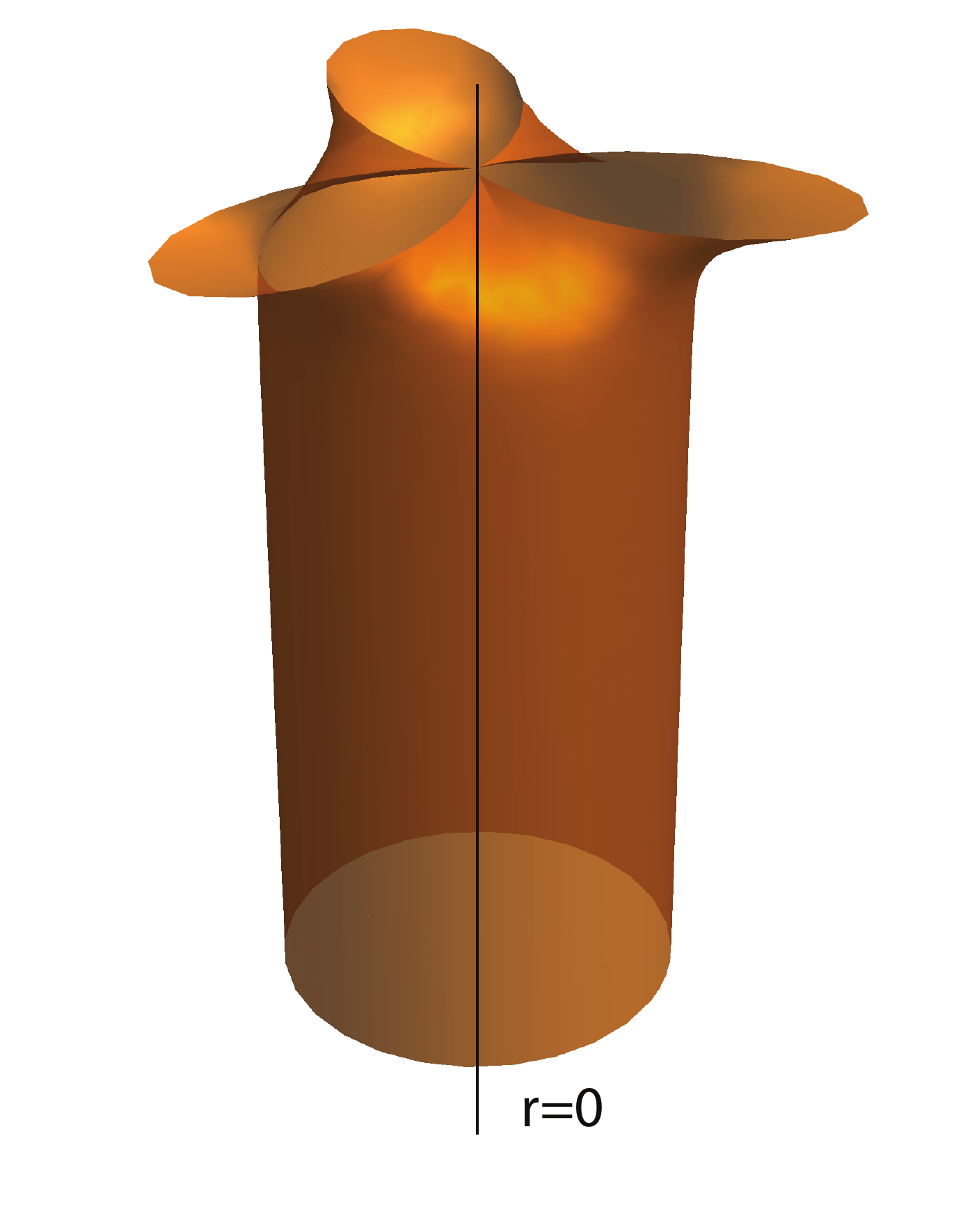}
\includegraphics[width=100mm]{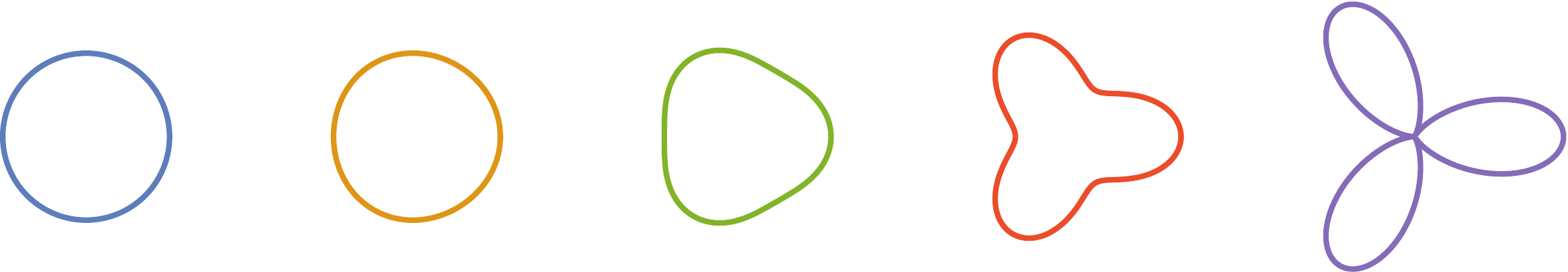}
\\{  $\lambda_{-\infty}=4.00$, $\lambda_0=4.00$, $\lambda_{\tau_\star/2}=4.00 $, $\lambda_{0.8 \tau_\star}=3.96$, $\lambda_{\tau_\star}=3.46 $ }
\caption{  Pictural representation and sections of the external horizon of the metric Eqs.[\ref{defUVbeta}, \ref{bMS}--\ref{VMS}] for the parameter choice : $r_+=4\,L$, $r_-=2\,L$, $\omega=0$, $Q=1$, $m^{(n>0)}_{(0,0)}=0$, $V^{(n>0)}_{(0,0)}=0$ excepted $V^{(3)}_{(0,0)}=V^{(-3)}_{(0,0)}=1/2$. The sections are considered at times : $\tau=-\infty$, $0$, $\tau_\star/2$, and $\tau_{0.8 \star}$  up to the moment $\tau_\star=0.46$ where it hits the singularity,  that becomes naked. The lengths $\lambda_\tau$ (in units $2\,\pi\,L$) of the horizon sections also are indicated.}\label{figure:HSN3}
\end{center}
\end{figure}

As we have shown in the preceding sections, all the asymptotic solutions may be obtained following a procedure involving solutions of the heat equation on a circle. There is a particular subset of solutions that can be written in closed form. Let us assume that for $p\geq 1$ all the
functions $f_{(q,p)}(\tau,\,\theta)$ and $a_{\phi(q,p)}(\tau,\,\theta)$ are vanishing.  The hierarchy of equations is almost all trivially satisfied. The only remaining non-zero functions are :
\begin{subequations}
\begin{align}  &f (\tau,\,\rho,\,\theta)=f_{(0,0)}(\tau,\,\theta)=-\frac 12\,\ln(Q)\qquad ,\label{fMS}\\
&\beta(\tau,\,\rho,\,\theta)=\ln(\rho/L)+\beta_{(0,0)}(\tau,\,\theta)=\ln(\rho/L)+ 2\,\ln(Q)\qquad ,\label{bMS}\\
&U(\tau,\,\rho,\,\theta)=L\,\frac \omega\rho\qquad ,\label{UMS}\\
&V(\tau,\,\rho,\,\theta)=-4\, \frac\rho L+ \sum_n V_{(0,0)}^{(n)}\,e^{i\,n\,\theta+Q^2\,n^2\,\tau}-\frac 14\,k_g\,\frac L\rho\,\Big( \sum_n m_{(0,0)}^{(n)}\,e^{i\,n\,\theta+Q^2\,n^2\,\tau}\Big)^2-\frac{L\,\omega^2}{Q^2\,\rho}\quad ,\label{VMS}\\
&A_\tau(\tau,\,\rho,\,\theta)=\frac{L^2}\rho\,\sum_n m_{(0,0)}^{(n)}\,e^{i\,n\,\theta+Q^2\,n^2\,\tau} +L\,\partial_\tau\,\lambda(\tau,\,\theta)\qquad ,\label{AtMS}\\
&A_\theta(\tau,\,\rho,\,\theta)=L\,c_\theta+L\,\partial_\theta\,\lambda(\tau,\,\theta)\qquad ,\label{AqMS}\\
&B_{\tau\theta}(\tau,\,\rho,\,\theta)=\frac{\omega\,L^3}{Q^2\,\rho}+\frac{k_g}{4}\, A_\tau(\tau,\,\rho,\,\theta)\,A_\theta(\tau,\,\rho,\,\theta)+L^2\,b_{\tau\theta(0,0)}(\tau,\,\theta)\qquad ,\label{BMS}\\
&H_{\tau\rho\theta}(\tau,\,\rho,\,\theta)=\frac{\omega\,L^3}{Q^2\,\rho^2}\qquad .\label{HMS}
\end{align}
\end{subequations}
Actually this solution was obtained from a direct integration of the field equations, under the working assumption $f_{(q,p>0)}(\tau,\,\theta)=0$, $a_{\theta(q,p>0)}(\tau,\,\theta)=0$ [see appendix \eqref{SpeSol}], but it could be guessed\footnote{Guessed, because we have limited the hierarchy  of equations at order $p=2$.} from the previous equations by noticing that eq. (\ref{eqa02}) implies $n_{(0,1)}=-\omega/Q^2$ which is (fortunately) compatible with eq. (\ref{eqf02}).\\
The Gauss curvature of the corresponding geometry (see eqs (\ref{defUVbeta})) is given by
\begin{align}R=-\frac{8}{Q^2\,\rho^2}-L\,\frac{\sum_n V_{(0,0)}^{(n)}\,e^{i\,n\,\theta+Q^2\,n^2\,\tau}}{Q^2\,\rho^3}+\frac{5\,L^2\,\omega^2}{2\,Q^4\,\rho^4}+k_g\,L^2\,\frac{\Big(\sum_nm_{(0,0)}^{(n)}\,e^{i\,n\,\theta+Q^2\,n^2\,\tau}\Big)^2}{2\,Q^2\,\rho^4}\qquad .\label{RMS}\end{align}
Solution (\ref{fMS}-\ref{HMS}) coincides  with the  black string (\ref{BSmet}--\ref{BSKRB}) when $\tau$ goes to $-\infty$. It diverges on the timelike singularity $\rho=0$ but also, when dynamical, on the surface $\tau=+\infty$, developing a "cosmological singularity". Let us remind that the more general solutions, discussed previously, present such singularities both on $\tau=+\infty$ and on $\tau=-\infty$ [See eq. (\ref{genXqp})]. These surfaces  correspond to Cauchy horizons of the stationary solution (\ref{BSmet}--\ref{BSKRB}). These singularities the dynamical solutions develop are in accordance  with the analysis of Chandrasekhar and Hartle in the frame of Reissner--Nordstr\"om black holes \cite{ChHa}.
\vspace{2mm}
As already mentioned, the dynamical configurations [Eqs (\ref{fMS}--\ref{HMS})] constitute deformations of black string geometries, which present two horizons [see Eqs(\ref{rprm})]. Accordingly we found interesting to examine the behaviour of the null surfaces (hereafter denoted by $\Sigma_\pm$)  coinciding with them asymptotically in the past.
 The evolution of these  surfaces is obtained by solving the partial differential equation :
\begin{align}\partial_\tau {\mathcal R}=\,&-\frac{k_g}{8} \,\frac{L^2}{\mathcal R}\Big( \sum_n m_{(0,0)}^{(n)}\,e^{i\,n\,\theta+Q^2\,n^2\,\tau}\Big)^2-\frac{\big(Q^2\,\partial_\theta{\mathcal R}-\omega\,L\big)^2}{2\,Q^2\,{\mathcal R}}\nonumber \\
&-2\,{\mathcal R}+\frac L2 \sum_n V_{(0,0)}^{(n)}\,e^{i\,n\,\theta+Q^2\,n^2\,\tau}
\label{EqH}\end{align}
with the asymptotic condition [see eq.(\ref{rprm})] :
$\lim_{\tau\rightarrow -\infty}{\mathcal R}=r_\pm$.
 \\A large degree of arbitrariness characterises  the geometry defined by the configuration [\ref{fMS}--\ref{HMS}]. It  is encoded in the function $V(\tau,\rho,\theta)$ (Eq.[\ref{VMS}]) that appears in the $g_{\tau\,\tau}$ component of the metric. This function $V(\tau,\rho,\theta)$ involves two infinite series of arbitrary coefficients\footnote{Only subjected to reality conditions :$V^{(n)}_{0,0)}=(V^{(-n)}_{0,0)})^\star$ and  $m^{(n)}_{0,0)}=(m^{(-n)}_{0,0)})^\star$)} : $V^{(n)}_{0,0)}$ and  $m^{(n)}_{0,0)}$. This arbitrariness make it difficult (if not impossible) to discuss the general structure of  these null surfaces for $\tau>0$. Of course, for $\tau\rightarrow -\infty$ only   \begin{align}
V^{(0)}_{0,0)}=4\frac{(r_++r_-)}L\quad \text{and}\quad  m^{(0)}_{0,0)}=\frac 4{\sqrt{k_g}}\sqrt{\frac{r_+\,r_-}{L^2}-\frac{\omega^2 }{4\,Q^2}}
\end{align}
 contribute and we recover the ``Kerr-like horizon structure'' discussed in ref. \cite{Detournay:2005fz}.
\\Preliminary numerical  integrations of eq. (\ref{EqH}), for  particular choices of the $V^{(n)}_{(0,0)}$ and $m^{(n)}_{(0,0)}$ arbitrary coefficients, illustrate some aspects of the  causal structure of the dynamical geometries. In particular they show that some  of the $\Sigma_\pm$  generators  may hit the singularity $\rho=0$,  while others run to infinity. We illustrate this behaviour by considering the special configuration where $\omega =0$ and where the equation :
\begin{align}
\partial_\theta V(\tau,\rho,\theta)=0\label{dV}
\end{align}
admits constant solutions $\theta=\theta_\star$, {\it i.e.} zeros independent of $\tau$ and $\rho$\footnote{For instance that will be the case if, for $n\neq 0$,  the arbitrary coefficients $V^{(n)}_{(0,0)}$ and $m^{(n)}_{(0,0)}$ are all real ({\it resp.} purely imaginary). In this case the series occurring in the expression of $V(\tau,\rho,\theta)$ only involves functions $\cos(n\,\theta)$ ({\it resp.} $\sin(n\,\theta)$) and we obtain $\theta_\star=0$ and $\pi$ ({\it resp.}  $\theta_\star=\pm \pi/2$). If moreover, for $n\neq 0$, the coefficients $V^{(n)}_{(0,0)}$ and $m^{(n)}_{(0,0)}$ are non-zero only for one value of $n$ than Eq.(\ref{dV}) will admit $2\,n$ solutions, equally spaced by $\pi/n$.}. In such case the curves $\{\theta=\theta_\star,\ \rho=\rho_\star(\tau)$ where $\rho_\star(\tau)$ is the solution of the first order  (asymptotic) Cauchy problem :
\begin{subequations}
\begin{align}
\frac {d\rho}{d\tau}&=\frac L 2\,V(\tau,\rho,\theta_\star)\qquad,\\
&=-  2  \Big(\rho -\big(r_++r_-+L/4\sum_{n\neq 0}V_{(0,0)}^{(n)}\,e^{i\,n\,\theta_\star+Q^2\,n^2\,\tau}\big)\nonumber\\
&+\frac{k_g}\rho\big(\sqrt{ {r_+\,r_-}/{k_g}}+ L/4\sum_{n\neq 0}m_{(0,0)}^{(n)}\,e^{i\,n\,\theta_\star+Q^2\,n^2\,\tau}\big)^2\Big)\qquad,\label{hgen}\\
\quad \rho_\star(-\infty)&=r_+\quad (\text {\it resp.}\ \rho_\star(-\infty)=r_-)\qquad ,
\end{align}
\end{subequations}
are generators of the external ({\it resp.} internal) horizon.
\\To go further let us specialise some more the arbitrariness  of the geometry we examine by fixing, for $n\neq 0$,  all the $m_{(0,0)}^{(n)}$ coefficients to zero.  Equation (\ref{hgen}) reduces to :
\begin{align}
&\frac{d\rho}{d\tau}=-2\frac {(\rho-r_+)(\rho-r_-)}\rho+\frac L2\,\sum_{n\neq0} V_{(0,0)}^{(n)}e^{i\,n\,\theta_\star+Q^2\,n^2\,\tau}\label{Hgen}\qquad .
\end{align}
For large positive value of $\tau$, it is the values of 
\begin{align}\kappa_\star(u)=\sum_{n\neq0} V_{(0,0)}^{(n)}e^{i\,n\,\theta_\star+Q^2\,n^2\,\tau}\label{kappa}
\end{align}
 that drive the behaviour of the function $\rho_\star(\tau)$. If $\kappa_\star(\tau)$ grows to $+\infty$ like $\exp(k^2\, \theta)$ ($k\in\mathbb N$), $\rho_\star(\tau)$ also grows as $\exp(k^2\, \theta)$. If on the other hand $\kappa_\star(\tau)$ decreases exponentially  to $-\infty$, $\rho_\star(\tau)$ also will decrease and reach zero for a finite value $\tau_\star$ of $\tau$.
For illustrative purposes let us choose  $\sum_{n\neq0} V_{(0,0)}^{(n)}e^{i\,n\,\theta+Q^2\,n^2\,\tau}=\cos(3\,\theta) \exp(9\,\tau)$, $r_+=4\,L$, $r_-=2\,L$, $\omega=0$ and $Q=1$. A numerical integration, whose results are displayed in Fig.(\ref{figure:HSN3}) shows the evolution of the external horizon from its asymptotically past circular structure of radius $r_+$ to a trefoil-like one. Three of the generators evolving at fix values of $\theta$ : those with $\theta_\star =\pi/3,\ \pi/2, \ 5\,\pi/3$ decrease to $\rho=0$, while those with $\theta_\star =0,\ 2\,\pi/2, \ 4\,\pi/3$ blow up to $+\infty$. 

\begin{figure}[ht]
\begin{center}
\includegraphics[width=100mm]{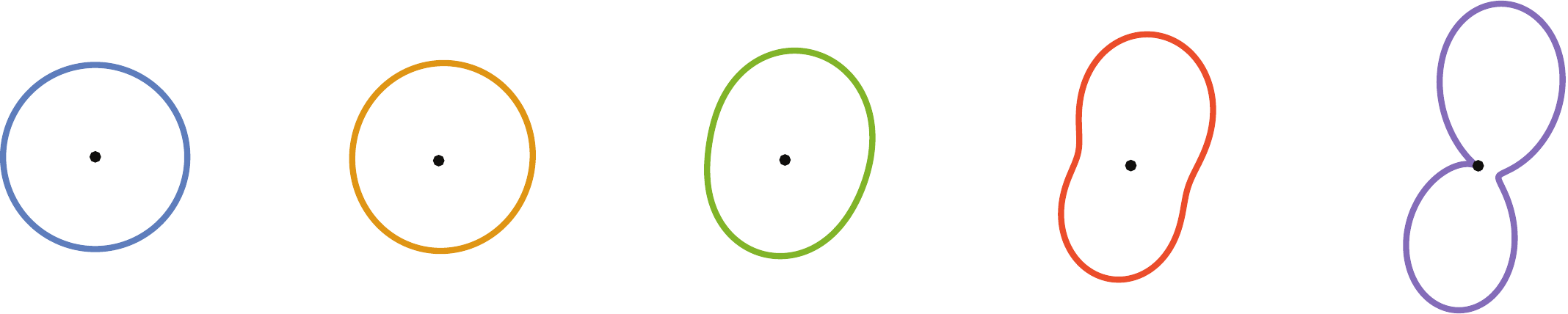}
\\{  $\lambda_{-\infty}=4.00$, $\lambda_0=4.00$, $\lambda_{\tau_\star/2}=3.98 $, $\lambda_{0.8 \tau_\star}=3.90$, $\lambda_{\tau_\star}=3.49 $ }
\caption{ Sections of the external horizon of the metric Eqs.[\ref{defUVbeta}, \ref{bMS}--\ref{VMS}] for the parameter choice : $r_+=4\,L$, $r_-=2\,L$, $Q=1$,  $\omega=5$, $\sum_{n\neq0} V_{(0,0)}^{(n)}e^{i\,n\,\theta+Q^2\,n^2\,\tau}=\sin(\theta)\exp(\tau)+\cos(2\theta)\exp(4\,\tau)$. 
. The sections are considered at times : $\tau=-\infty$, $0$, $\tau_\star/2$, and $\tau_{0.8 \star}$  up to the moment $\tau_\star=0.86$ where it hits the singularity (represented by black dots)   that becomes naked. The lengths $\lambda_\tau$ (in units $2\,\pi\,L$) of the horizon sections also are indicated.} 
\label{figure:HSN12w5}
\end{center}
\end{figure}

\newpage
The main purpose of this illustrative discussion is to show that the structure of the horizon of the black string is geometrically rich, far from being completely understood, and deserves extra studies. Before stopping it, let us show another sequence of horizon sections and their respective lengths [Fig. (\ref{figure:HSN12w5}) ] for a more general configuration, where $\omega\neq0$ and which involves both sine and cosine functions : $r_+=4\,L$, $r_-=2\,L$, $Q=1$,  $\omega=5$, $\sum_{n\neq0} V_{(0,0)}^{(n)}e^{i\,n\,\theta+Q^2\,n^2\,\tau}=\sin(\theta)\exp(\tau)+\cos(2\theta)\exp(4\,\tau)$. 
Here again we see that some horizon generators hit the singularity (despite the absence of symmetry) while others (apparently) go to infinity.
 \\ To make an end to these numerical considerations, let us mention that they also indicate that the length of the horizon diminishes when the time variable $\tau$ increases. This is not in contradiction with the usual theorem on black holes as here  the horizons are  past horizons. A simple way to reconcile the behaviour of the horizon with the thermodynamic law of entropy increase is to change $\tau$ into $-\tau$. But then the $\tau$ variable has to be interpreted as a advanced time  labelling ${\mathcal J}^-$ and not as a retarded time on ${\mathcal J}^+$.  The corresponding solutions describe configurations where initially divergent field modes vanish and  futur horizons appear, masking an initial naked singularity.

\section{Charges and asymptotic charges}\label{Charges}

Solutions of gauge invariant theories that admit gauge transformations leaving the field configuration invariant, i.e.  a set of reducibility parameters $\{\zeta\}$, allow to define elementary charges, and, if integrable, finite charges (see for example refs \cite{Barnich:2001jy,Barnich:2007bf,Compere:2018aar} and references therein). The elementary charges are given by integration on closed surfaces of co-dimension 2, here on closed curves $\mathcal C $, of the elementary surface charge density (\ref{NWk}), here : $k_{\{\zeta\}\rho}:=k_{\{\zeta\}}^{\mu\nu}\eta_{\mu\nu\rho}$. Those   that we consider hereafter, are specifyed by choosing the solutions of the linearised field equations, the definition of $k_\zeta$ requires, given by arbitrary variations of the constants occurring in the expressions of the background solutions. Let us notice that if $\{\xi^\mu,\,\Lambda_\alpha,\,\lambda\}$ constitute a set of reducibility parameters, they all transform covariantly with respect to diffeomorphisms but under the gauge transformations :
\begin{align}
A_\alpha&\mapsto A_\alpha+\partial_\alpha \psi\qquad ,\\
B_{\alpha\beta}&\mapsto B_{\alpha\beta}+\partial_{[\alpha}\Psi_{\beta]}-\frac{k_g}4\partial_{[\alpha}\psi A_{\beta]}\qquad .
\end{align}
according to the rules :
\begin{align}
\xi^\mu&\mapsto \xi^\mu\qquad,\\
\lambda&\mapsto \lambda-\xi^\alpha\partial_\alpha \psi+\lambda^c \qquad ,\\
\Lambda_\alpha&\mapsto \Lambda_\alpha+ \frac{k_g}4\big(2\,\psi\, \partial_\alpha\lambda- (\xi^\beta\partial_\beta\psi)\partial_\alpha \psi\big)-\xi^\beta\partial_\beta \Psi_\alpha+\Lambda^c_\alpha \qquad,
\end{align}
where $\lambda^c$ is an arbitrary constant (a closed zero-form) and $\Lambda^c_\alpha$ the components of an arbitrary closed one-form.
Thus, without loss of generality, from now we may assume that the functions $b_{\phi\,u(0,0)}(u,\,\phi)$ in eq. (\ref{expB}), $b_{\tau\,\theta(0,0)}(\tau,\,\phi)$ in eq. (\ref{BMS}),  $\lambda(u,\,\phi)$ in eq.(\ref{A00}) and $\lambda(\tau,\,\theta)$ in eqs (\ref{AtMS}, \ref{AqMS}) to be zero, respectively.\par
The black string configuration Eqs~(\ref{BSmet}--\ref{BSKRB}) admits such parameters $\{\zeta\}:=\{\xi^\mu,\,\Lambda_\alpha,\,\lambda\}$ constituted by the  Killing vector fields $\partial_\tau$ and $\partial_\theta$ of the metric, ensuring that $\delta g_{\mu\nu} =0 $ and leaving the dilaton invariant $\delta \Phi=0$, and gauge parameters $\Lambda_\alpha$ (given by eq. (\ref{reducLambda}) and $\lambda$  (constant) such that $\delta A_{\alpha} =0 $ and $\delta B_{\alpha\,\beta}=0$  respectively (see eqs (\ref{metgaugeT}--\ref{PhigaugeT})). They lead to elementary charges that all appear to be integrable.\\
Its Killing horizons, located at $r=r_+$ and $r=r_-$ [see eq. (\ref{rprm})], are generated by the vector fields :
\begin{equation}
\gamma_\pm=\partial_\tau-\omega\,\frac L{r_\pm}\partial_\theta\qquad.
\end{equation}
This structure allows us to obtain a first law relating variation of energy, angular momentum and charges.  It requires that the gauge parameter are reducibility parameters, i.e.  when, evaluated on the black string background, $\xi^\mu$ are the components of the Killing vector fields, $\lambda$ is constant and $\Lambda=dF(\tau,\,\rho,\,\theta)+\Lambda_\theta\,d\theta$. Associated to them are conserved charges: energy, angular momentum and abelian (\ref{QMax}) and Kalb-Ramond (\ref{QKR}) charges respectively.  A standard calculation, based on the surface charge density provided in Appendix (\ref{AppSurfCh}) leads to :
\begin{align}\delta {\mathcal E}=\frac 1{8\,\pi\,G}\kappa_\pm\,\delta A_\pm-\omega\,\frac L{r_\pm}\,\delta J+\Phi_{Max}\,\delta {\mathcal Q}_{Max}+\Phi_{KR}\,\delta {\mathcal Q}_{KR}\qquad .\label{firstlaw}\end{align}
Here ${\mathcal E}=(r_++r_-)/(2\,G)$ is the energy, $
\kappa_\pm=\sqrt{-1/2\,\gamma^\mu_{\pm;\nu}\,\gamma^{\nu}_{\pm;\mu}}=2\,( {r_\pm-r_\mp})/{r_\pm}$ 
is the surface gravity and  $A_\pm=2\,\pi\,r_\pm$ the length of the outer horizon ($r=r_+$) or the inner horizon ($r=r_-$). Let us emphasize the contribution of the topological term introduced in the angular component of the abelian potential (\ref{AqMS})~:
\begin{equation}
J:=-\frac L{8\,G}\Big(\frac \omega{Q^2}+2\,c_\theta\,\sqrt{k_g}\sqrt{\frac{r_+\,r_-}{L^2}-\frac{\omega^2}{4\,Q^2}}-\frac{k_g}4\,c_\theta^2\,\omega\Big)\qquad .
\end{equation}
The charges are given by eqs (\ref{QMax}, \ref{QKR}) :
\begin{equation}
{\mathcal Q}_{Max}= \frac 1{8\,G}\Big(2\,\sqrt{k_g}\sqrt{\frac{r_+\,r_-}{L^2}-\frac{\omega^2}{4\,Q^2}}-\frac{k_g}2\,c_\theta\,\omega\Big)\quad ,\quad
{\mathcal Q}_{KR}=-\frac\omega{8\,G\,L}\qquad .
\end{equation}
The N\oe ther-Wald surface charge expression (\ref{NWsurfch})  provides those of the associated potentials. For the abelian fields, it takes its usual expression :
\begin{equation}
\Phi^\pm_{Max}= \gamma_\pm^\mu\,A_\mu\Big\vert_{r_\pm}= \frac {L^2}{r_\pm}\Big(\frac 4{\sqrt{k_g}}\sqrt{\frac{r_+\,r_-}{L^2}-\frac{\omega^2}{4\,Q^2}}-c_\theta\,\omega\Big)\qquad .
\end{equation}
However, due to a more complicated gauge transformation rule  of the $B$ field  (\ref{MaxgaugeT}), the corresponding potential linked to the Kalb-Ramond field reads :
\begin{equation}
\Phi^\pm_{KR}= \gamma_\pm^\mu\big(B_{\mu\,\theta}-\frac {k_g}4\,A_\mu\,A_\theta) \Big\vert_{r_\pm}= -\frac {L^3}{r_\pm}\Big( \frac\omega {Q^2}+2\,c_\theta\, {\sqrt{k_g}}\sqrt{\frac{r_+\,r_-}{L^2}-\frac{\omega^2}{4\,Q^2}}-
\frac{k_g}4\,c^2_\theta\,\omega\Big)\quad .
\end{equation}

Remarkably   all these relations remain valid in the framework of the solution described by eqs (\ref{fMS}--\ref{HMS}) if we lightly restrict the phase space by fixing $c_\theta$, i.e. restricting the solutions of the linearised equations by imposing \begin{equation}\delta c_\theta =0\qquad.\label{phsprestdcq}\end{equation} Indeed while the vector fields $\partial_\tau$ and $\partial_\theta$ are no longer the generators of isometries preserving the matter fields (even asymptotically), they nevertheless constitute symplectic symmetries \cite{Comp_re_2016,Comp_re_2015} : the surface charge density constructed with them and the fields eqs (\ref{fMS}--\ref{HMS}) is conserved.


We also may look for asymptotic symmetries. We restrict ourself to the search of such symmetries leading to asymptotic conserved charges.
Such asymptotic charges that may be linked to the field configurations described in the previous section rest on  the existence of asymptotic reducibility parameters i.e.  gauge transformations that preserve the asymptotic structure of the field configurations. In our case the asymptotic field behaviour is defined by eqs (\ref{ordres}, \ref{ordref}) and the Bondi gauge conditions eqs (\ref{gaugecond}). 
Such gauge transformations are determined as follows (see ref. \cite{Barnich:2015jua}, appendix A.1).\\
Imposing on the metric the Bondi gauge condition, we have to maintain the following three conditions~:
\begin{align}g_{r\,r}=0\qquad,\qquad g_{r\,\phi}=0\qquad,\qquad g_{\phi\,\phi}=r^2\qquad,\end{align}
i.e.  in terms of the infinitesimal diffeomorphisms (\ref{metgaugeT}) to impose that their generators $\xi^\alpha$   satisfy the three conditions 
\begin{align} {\mathcal L}_\xi g_{r\,r}&=0\qquad\mbox{which implies} \qquad \partial_r\xi^u=0{\nonumber  }\\
&\phantom{=0\qquad } \mbox{i.e.}\quad \xi^u=X(u,\,\phi)\qquad ,\\
 {\mathcal L}_\xi g_{r\,\phi}&=0\qquad\mbox{which implies} \qquad \partial_r\xi^\phi-\frac{L\,e^{\beta}}{r^2}\partial_\phi X(u,\,\phi)=0{\nonumber  }\\
&\phantom{=0\qquad } \mbox{i.e.}\quad \xi^\phi=Y(u,\,\phi)+L\, \partial_\phi X(u,\,\phi)\,\int \frac{ e^{\beta}}{r^2}dr\\
&\phantom{=0\qquad   \mbox{i.e.}\quad \xi^\theta}= \partial_\phi X(u,\,\phi)\, e^{\beta_{(0,0)}}\ln(r/L)+{\mathcal O}(1)\qquad ,\label{putxiphi}\\
 {\mathcal L}_\xi g_{u\,\phi}&=0\qquad\mbox{which implies} \qquad \xi^r=-r\big(U\,\partial_\phi \xi^u+\partial_\phi \xi^\phi)\qquad .
\end{align}
Moreover, imposing that $g_{u\,r}$ remains at order $r$ :
\begin{align}{\mathcal L}_\xi g_{u\,r}={\mathcal O}(r)\qquad ,\end{align}
leads to an equation dictated by  the term of order ${\mathcal O}(\ln(r/L))$ in eq. \ref{putxiphi}:
\begin{align}\partial_\phi \xi^u\,\partial_\phi e^{\beta_{(0,0)}(u,\,\phi)}=2\,\partial_\phi \big(\partial_\phi \xi^u\,e^{\beta_{(0,0)}(u,\,\phi)}\big)\end{align}
whose solution is
\begin{align}\partial_\phi \xi^u=\Delta(u)\,e^{-\ft 12\beta_{(0,0)}(u,\,\phi)}\qquad .\end{align}
But, for $\xi^u$ to be periodic we have to require that $\Delta(u)=0$. As a consequence we obtain :
\begin{align}\xi^u=X(u)\qquad,\qquad \xi^r=-r\,\partial_\phi Y(u,\,\phi) \qquad,\qquad \xi^\phi= Y(u,\,\phi)\qquad.\label{xias}\end{align}
The remaining assumption on the asymptotic behaviour of the metric components~: $g_{u\,u}={\mathcal O}(r^2)$ and $g_{u\,\phi}={\mathcal O}(r^2)$ and the dilaton field does not imply any further condition. It is worthwhile to notice that in particular we obtain :
\begin{align}{\mathcal L}_\xi \Phi={\mathcal O}(1)\end{align}
in accordance with the   ansatz  eq. (\ref{fdef}).

The Bondi gauge conditions eqs (\ref{gaugecond}) on the abelian and the Kalb-Ramond fields are preserved by the infinitesimal diffeomorphisms defined by the vector field eq. (\ref{xias}). The residual gauge transformations Eqs(\ref{MaxgaugeT}--\ref{KRgaugeT}) that preserve conditions  (\ref{gaugecond})  are :\begin{align}\lambda=\ell(u,\,\phi)\qquad,\qquad \Lambda =dL(u,\,r,\,\phi) +P_\phi(u,\,\phi)\,d\phi\qquad.\label{BondiGaugeCond}\end{align}
This expression of the one-form $\Lambda$ as written is not canonical. It can as well be written in terms of three component functions :
\begin{align} &\Lambda =K_u(r,\,u,\,\phi)\,du+K_r(r,\,u,\,\phi)\,dr +K_\phi(r,\,u,\,\phi)\,d\phi\end{align}
linked by the two integrability conditions :
\begin{align}\quad \partial_r K_u(r,\,u,\,\phi)-\partial_u K_r(r,\,u,\,\phi)=0,\quad\mbox{and}\quad\partial_r K_\phi(r,\,u,\,\phi)-\partial_\phi K_r(r,\,u,\,\phi)=0\label{BondiGaugeCond2}\end{align}

The asymptotic symmetry algebra is thus expressed in terms of triples $s=(\xi,\,\ell,\,\Lambda)$ of vectors $ { \xi}$ of the form eq. (\ref{xias}), function $\lambda$ of two variables and one-form obeying the conditions eq. (\ref{BondiGaugeCond2}). 

The algebra they satisfy is given, in terms of usual Lie derivatives $\cal L$ by :
\begin{align} &[(\xi_1,\ell_1,\Lambda_1),(\xi_2,\ell_2,\Lambda_2)]=(\xi,\ell,\Lambda)\qquad\mbox{with}\\
&\xi={\cal L}_{\xi_2}\xi_1-{\cal L}_{\xi_1}\xi_2\quad ,\quad  \ell={\cal L}_{\xi_2}\ell_1-{\cal L}_{\xi_1}\ell_2\quad ,\quad 
\Lambda={\cal L}_{\xi_2}\Lambda_1-{\cal L}_{\xi_1}\Lambda_2+\frac{k_g}4\Big(\ell_1\,d\ell_2-\ell_2\,d\ell_1\Big)\qquad .
\end{align}
In particular the asymptotic Killing vector fields realise the semidirect product of two Wit algebras.

But this is not the end of the computation of asymptotic reducibility parameters. To simplify the analysis let us fix the background by imposing the extra conditions eqs (\ref{simpcond}). To obtain conserved\footnote{Non conserved asymptotic charges and their algebra are considered in refs \cite{Barnich:2015jua,PhDCZ,Detournay:2018cbf}; see also appendix (\ref{AppSurfCh}). } asymptotic charges, remembering that we are considering polyhomogeneous asymptotic expansions, we also have to require that :
\begin{subequations}
\begin{align}
\partial_\alpha \big(\sqrt{-g}\,k^{\alpha\,\tau}\big)=&\mathcal O\Big(\frac{\ln[\rho/L]^{n_\tau}}{\rho^2}\Big)\label{divcondu}\qquad , \\
\partial_\alpha \big(\sqrt{-g}\,k^{\alpha\,\rho}\big)=&\mathcal O\Big(\frac{\ln[\rho/L]^{n_\rho}}{\rho}\Big)\qquad ,\label{divcondr}\\
 \partial_\alpha\big( \sqrt{-g}\,k^{\alpha\,\theta}\big)=&\mathcal O\Big(\frac{\ln[\rho/L]^{n_\theta}}{\rho^2}\Big)\qquad , \label{divcondphi}
\end{align}
\end{subequations}
where $n_\tau$, $n_\rho$ and $n_\theta$ are all three non-negative integers. These conditions insure both that the elementary charges defined by the integral (\ref{elemch}) are independent of the shapes of the loops on which the various integrals are evaluated and that they are conserved in time.
Imposing that the stationary black string configuration belongs to the phase space, a straightforward calculation then leads to the conclusion that only constant reducibility parameters :
\begin{equation}
\xi^\tau=X_h\quad,\quad\xi^\theta=Y_j\quad,\quad \lambda=\ell\quad,\quad \Lambda=\Lambda_\theta \,d\theta
\end{equation}
constitute asymptotic reducibility parameters. Indeed we obtain that for the stationary black string configuration eqs (\ref{metrpm}, \ref{FHrpm}) the   relevant parts of the components of the divergence of the charge density are given by  :
\begin{subequations}
\begin{align}
\partial_\alpha \big(\sqrt{-g}\,k^{\alpha\,\tau}\big)=& 0\qquad,\\
\partial_\alpha \big(\sqrt{-g}\,k^{\alpha\,\rho}\big)=& \frac1{16\,\pi\,G}\left(\rho\Big(\big(Q^2\,Y'''(\tau,\,\theta)-\dot Y'(\tau,\,\theta)\big)\frac{\delta Q}Q+\frac{k_g}4\,Q^2\big(\dot\ell'(\tau,\,\theta)+c_\theta\,\dot Y'(\tau,\,\theta)\big)\,\delta c_\theta\Big)\right.\nonumber\\
&+8\,G\Big(\dot X(\tau)\,\delta\mathcal E- \dot Y(\tau,\,\theta)\,\delta  J +L\,\dot \ell(\tau,\,\theta)\,\delta \mathcal Q_{Max}-L^2\,\dot K_\theta(\tau,\,\theta)\,\delta \mathcal Q_{KR}\Big)\nonumber\\
&+{\sqrt{k_g}}\,L\,Q^2\,Y''(\tau,\,\theta)\,\left(2\,\sqrt{\frac{r_+\,r_-}{L^2}-\frac{\omega^2}{4\,Q^2}}\,\delta c_\theta-c_\theta\,\delta\left( \sqrt{\frac{r_+\,r_-}{L^2}-\frac{\omega^2}{4\,Q^2}}\right)\right) \nonumber\\
&\left.-{\sqrt{k_g}}\,L\,Q^2\,\ell''(\tau,\,\theta)\,\delta \left(\sqrt{\frac{r_+\,r_-}{L^2}-\frac{\omega^2}{4\,Q^2}}\right)\right)+  \mathcal O[ 1/ r]\qquad ,\\
\partial_\alpha \big(\sqrt{-g}\,k^{\alpha\,\theta}\big)=&  \frac1{16\,\pi\,G}\left(\big(\dot Y(\tau,\,\theta)-Q^2\,Y''(\tau,\,\theta)\big)\frac{\delta Q}Q-\frac{k_g}4\,Q^2\big(\dot\ell(\tau,\,\theta)+c_\theta\,\dot Y(\tau,\,\theta)\big)\,\delta c_\theta\nonumber\right.\\
&\left.-\frac L \rho \dot X(\tau)\,\delta \omega\right) \qquad .
\end{align}
\end{subequations}
 However, if we restrict the phase space by fixing $Q$ and $c_\theta$ we obtain a larger set of reducibility parameters :
\begin{align}
&\xi^\tau=X_h\ ,\ \xi^\theta=Y[\tau,\,\theta]\quad (arbitrary) \ ,\ \lambda=-c_\theta\,Y[\tau,\,\theta]+\sum_n\ell^{(n)}\,e^{i\,n\,\theta-Q^2\,\frac{n^2}2\,\tau}\ ,\nonumber\\& \Lambda=\left(-\frac{4)k_g\,c_\theta^2\,Q^2}{Q^2}\,Y[\tau,\,\theta]+\Lambda_\theta\right)d\theta\quad .
\end{align}
But if we only impose $c_\theta$ constant on the all phase space, the reducibility parameter set becomes more restricted. Instead to have $\xi^\theta$ arbitrary it must be taken constant : $\xi^\theta= Y_j$ constant to define conserved asymptotic charges.

Let us now consider the conditions eqs (\ref{divcondu}--\ref{divcondphi}) applied  on the more general asymptotic solutions obtained in section (\ref{IntasympEqs}) we are led  to restrict the covariant phase space as follows. To satisfy the first of the conditions eq. (\ref{divcondu}) we have to require  : 
\begin{align}
f_{(0,1)}(\tau,\,\theta)=f^{(0)}_{(0,1)}\quad,\quad a_{\theta(0,1)}(\tau,\,\theta)=a^{(0)}_{\theta(0,1)}+4 \,\tau\,a^{(0)}_{\theta(1,1)}\quad,\quad a^{(0)}_{\theta(1,1)}\,\delta c_\theta=0\qquad.\label{cda11dcq}
\end{align}
For instance, if the conditions (\ref{cda11dcq}) are not satisfyed, we obtain that, at fixed value of $\tau=\tau_\star$, the integral of the surface charge density, on a circle of large radius $\rho=r_\star$,  behaves as :
 \begin{align}
 \int_0^{2\,\pi} k_{\{X_h,Y_j,\ell,\Lambda_\theta\}}^{\tau,\rho}[\tau_\star,r_\star,\theta]d\theta=&4\,\pi\,X_h\,L\,Q^2\,k_g\,\delta c_\theta\,a_{\theta(1,1)}\, \Big(\ln(r_\star/L)+4\,\tau_\star\Big)\nonumber\\&+\mbox{$\tau_\star$ and $r_\star$ independent terms} \qquad .
\end{align}

The three conditions eq. (\ref{cda11dcq}), combined with eqs (\ref{solm00}, \ref{soln01}, \ref{solV00}) imply that :
\begin{align}
&\delta \Big(Q^2\big(\dot V_{(0,0)}(\tau,\,\theta)+Q^2\,V''_{(0,0)}(\tau,\,\theta)\big)\Big)=0\qquad ,
\end{align}
and similar relations for $m_{(0,0)}(\tau,\,\theta)$ and $n'_{(0,1)}(\tau,\,\theta)$.
 Using them, we see that to verify condition (\ref{divcondr}), we  also have to  impose that :
\begin{align}
n'_{(0,1)}(\tau,\,\theta)\,\delta Q =0\qquad,\qquad \dot m_{(0,0)}(\tau,\,\theta)\,\delta c_\theta =0\qquad .
\end{align}
Then the third condition (\ref{divcondphi}) implies no more restriction.

 Thus in order to obtain conserved, well defined elementary charges, we have, together with the first two   conditions (\ref{cda11dcq}), four options to define covariant phase spaces: to assume $c_\theta$ fixed or  $a_{\theta(1,1)}^{(0)}=0$ and $m_{(0,0)}(\tau,\,\theta)=m_{( 0,0)}^{(0)}$ and $Q$ fixed or $n_{(0,1)}(\tau,\,\theta)=n_{( 0,1)}^{(0)}$.   According to the choice, we obtain as elementary charges :
\begin{align}
\delta \mathcal E&=\frac{1}{8\,G}\Big(\delta V^{(0)}_{(0,0)}-16\frac{f_{(0,1)}^{(0)}}Q\,\delta Q+ 2\,k_g\,Q^2\,a_{\theta(0,1)}^{(0)}\delta c_\theta\big)\qquad,\label{genEldE}\\
\delta J&=\frac{1}{8\,G}\delta\Big(n_{(0,1)}^{(0)}-\frac 12\,k_g\,c_\theta\,m_{(0,0)}^{(0)}-\frac 14\,\omega\,c^2_{\theta}\Big)\qquad ,\\
\delta \mathcal Q_{Max}&=\frac{1}{16\,G}k_g\,\delta\Big(m_{(0,0)}^{(0)}-\omega\,c_{\theta}\Big)\qquad ,\\
\delta \mathcal Q_{KR}&=-\frac 1{8\,G\,L}\delta\omega\qquad .
\end{align}
In the above expressions we have to set $\delta c_\theta=0$ if we choose the phase space sector to have $a_{(1,1)}^{(0)}\neq 0$. All these elementary charge remain integrable, excepted for the energy (the one linked to the asymptotically reducibility parameter $X_h$, corresponding to $\tau$ translation), unless we assume $ Q$ and $c_\theta$ fixed, or more generally appropriate special $Q$ and $c_\theta$ dependences for the $f_{(0,1)}^{(0)}$ and $a_{\theta(0,1)}^{(0)}$ integration constants.
\section{Conclusions}
We have obtained a (formal) generalisation of our previous results of black string configurations, established in ref. \cite{Detournay:2005fz}, that describes a general solution of the Einstein-dilaton-abelian-Kalb-Ramond field equations. This dynamical system depends on two degrees of freedom. The solution we obtain depends on an infinite set of arbitrary functions of $\tau$ and $\theta$ that we may reinterpret as Cauchy data  expressed as series involving power of $\rho$ and $\log [\rho/L]$. Surprisingly we have obtained a closed form solution of the problem without isometry--gauge invariance : a solution describing a dynamical black string configuration. Nevertheless this solution exhibits a hidden symplectic symmetry that  allows us to define for it charges which are similar to that of the stationary black string configuration. We also have shown, that restricting some of the arbitrary functions occurring in the expansion of the general asymptotic solution to constants, we may still define four conserved elementary charges. Three of these charges are integrable. Only the energy  requires some extra conditions to be defined.
Actually, that the abelian and Kalb-Ramond charges always exist results from eqs (\ref{QMax}, \ref{QKR}). Less expected, on the contrary to what is necessary for the asymptotic energy,  the asymptotic angular momentum didn't require extra conditions to be defined, .

A few other points deserve attention. \\
Firstly : the algebra of the asymptotic reducibility parameters may seem frustrating: in general it coincides with the four-dimensional algebra of isometry--gauge transformations preserving the stationary configuration. 
\\Secondly : the black string configuration is proved to be unstable, as conjectured by Horne and Horowitz on the basis of to the work of Chandrasekhar and Hartle.   Of course the link between these instabilities and their CFT interpretation in terms of deformations by marginal operators of the underlying sigma-model will require extra work.  \\
Thirdly : the subdominant terms occurring in the asymptotic expansion of the general solution have to be restricted in order to obtain conserved charges or even only finite non conserved charges [see appendix (\ref{AppSurfCh}) ]\\
Fourthly  : the asymptotic solutions are mainly determined by the two functions eqs (\ref{asf}, \ref{asAphi}). Their expansion in terms of inverse powers or $r$ and power of $\ln[r/L]$ involves arbitrary (complex) constants determining the coefficients $f_{(q,p)}(u,\,\phi)$ and $a_{(q,p)}(u,\,\phi)$ and their time derivatives. Roughly speaking they corresponds to the initial data of these two degrees of freedom and thus, for each choice of them define inequivalent physical configurations.
 
\appendix
 \section {Geometrical description of the asymptotic geometry\label{GeomEmb}}

 On a four dimensional Minkowski space, with metric :
\begin{align}dS^2= dU\,dV+dX^2+dY^2\end{align} we have to consider the half-cones with equation :\begin{align}U>0\quad,\quad p^2\,U\,V-q^2(X^2+Y^2)=0\quad,\quad p^2+q^2=\frac 14\quad p,\ q\in ]0,\ft12[\quad , \label{cone2}\end{align}
parametrised by :
\begin{align}U=q\,Q\,L\,(\rho/L)^{\ft{2\,q-1}{2\,q}}e^{-\ft 2 q\tau}\ ,\  V=q\,Q\,L\,(\rho/L)^{\ft{2\,q+1}{2\,q}}e^{\ft 2 q\tau}\ ,\  X=p\,Q\,\rho \cos(\ft {\theta} {p\,Q})\ ,\  Y=p\,Q\,\rho \sin(\ft {\theta} {p\,Q})\ .\end{align}
The $p$   and $q$ parameters have to satisfy   the  rationality condition ($T$ being the period of the cyclic coordinate $\theta$, choosen to be $2\,\pi$ in the main text):
\begin{align}\frac T{2\,\pi\, p\,Q}\in {\mathbb Q}\qquad ,\end{align}
in order for the cones given by eq. (\ref{cone2}) to constitute  finite coverings of the manifold we are looking for. The induced metric on this surface is given by eq. (\ref{dsas}).
\section{Special solution}
\label{SpeSol}
We start from the special field configuration ansatz :
 \begin{align}{&\Phi=-\frac 12\ln\left(\frac {Q\,\rho}{ L }\right)\\&A_\mu=\left\{A_\tau(\tau,\,\rho,\,\theta),0,L\,c_\theta)\right\}}\end{align}
but the general metric components (in Bondi gauge) :
 \begin{align} g_{\mu\nu}=\left(
 \begin{array}{ccc}
  g_{\tau\,\tau}(\tau,\,\rho,\,\theta) &  g_{\tau\,\rho}(\tau,\,\rho,\,\theta)  &   g_{\tau\,\theta}(\tau,\,\rho,\,\theta)\\
 g_{\tau\,\rho}(\tau,\,\rho,\,\theta) & 0  &   0\\
   g_{\tau\,\theta}(\tau,\,\rho,\,\theta)&  0 &  \rho^2 
 \end{array}\label{metgenBondi}
 \right) \qquad  .
 \end{align}
The $H_{\mu \nu \rho}$ field components are given by eq. \eqref{expH} :
\begin{align}H_{\tau\,\rho\,\theta}=\frac {{{L}}^3 \omega\, \vert g_{\tau\,\rho}(\tau,\,\rho,\,\theta)\vert}{r^3}\qquad .\end{align}
 The Einstein equation  $\mathcal E_{\rho\,\rho}=0$ [see eqs \ref{EqG}] implies that :
\begin{align}{ g_{\tau\,\rho}(\tau,\,\rho,\,\theta)=\frac r{L} \gamma_{\tau\,\rho}(\tau,\,\theta)}\qquad ,\end{align}
while $\mathcal E_{\rho\,\theta}=0$ provides :
\begin{align} 
g_{\tau\,\theta}(\tau,\,\rho,\,\theta)=L\,\rho\left(\gamma^{(1)}_{\tau\,\theta}(\tau,\,\theta)+( r/{L})\left[\gamma^{(2)}_{\tau\,\theta}(\tau,\,\theta)+\gamma^{\prime}_{\tau\,\rho}(\tau,\,\theta)\Big(1-\ln\left( r/{L}\right)\Big)\right]\right)\qquad  .
\end{align}
The abelian field equation  $\mathcal J^\tau=0$ [see eqs (\ref{EqA})] leads to the expected radial dependence of the potential~:
\begin{align} A_\tau(\tau,\,\rho,\,\theta)=\frac {L^2}\rho m(\tau,\,\theta) +L\,a_{\tau(0,0)}(\tau\,\theta)\qquad .
\end{align}
The equation $\mathcal J^\theta=0$  involves three terms, that are respectively proportional to  $\rho^{-2}$, $\rho^{-3}$  and  $\rho^{-4}$. The first two  imply that :
\begin{align}a_{\tau(0,0)}(\tau\,\theta)=&\lambda(\tau)\qquad,\\  \gamma_{\tau\,\rho}=&-b(\tau)\qquad ,\end{align}
where $\lambda(\tau)$ and $b(\tau)$ are arbitrary functions. Thanks to an appropriate choice of  gauge  we may assume  that $\lambda(\tau)=0$ and by redefining the $\tau$ variable that  $b(\tau)=Q^2$. With this choice of the $\tau$ variable the term of order $\rho^{-4}$ in the equation $\mathcal J^\theta=0$ fixes $\gamma^{(1)}_{\tau\,\theta}$ :
\begin{align}{\gamma^{(1)}_{\tau\,\theta}= \omega\qquad      .}\end{align}
Let us now turn to the dilaton-field equation \eqref{EqPhi}. It implies that :
\begin{align} g_{\tau\,\tau}=L\,\rho\, \gamma_{\tau\,\tau}(\tau,\,\theta)-\frac {k_g }{4}\,L^2\,Q^2\,m(\tau,\,\theta)^2+\rho^2\Big(\gamma^{(2)}_{\tau\,\theta}(\tau,\,\theta)^2+\gamma_{\tau\,\theta}^{(2)\prime}(\tau,\,\theta)-4\Big)\qquad,\end{align}
  an expression that the gravitational equation $Eqg_{v\,r}=0$ simplify as it requires that :
 \begin{align}{\gamma^{(2)}_{\tau\,\theta}(\tau,\,\theta)=a(\tau)=:\dot A(\tau)\qquad  .}\end{align}
  Here we have introduce a primitive $A(\tau)$ of the arbitrary function $a(v)$.\\
  The last abelian field equation $\mathcal J^\rho=0$ leads us to the evolution equation of the $m(\tau,\,\theta)$ function :
  \begin{align}  \dot m(\tau,\,\theta)=- Q^2\,m''(\tau,\,\theta)+a(\tau) \, m' (\tau,\,\theta)\label{evolm} \end{align}
  whose general solution reads
\begin{align} m(\tau,\,\theta)=\sum_n m^{(n)}_{(0,0)} \,e^{i \,n(A(\tau)+\theta)+n^2 \,Q^2\,\tau}\quad \mbox{with} \quad m^{(-n)}_{(0,0)}=(m^{(n)}_{(0,0)})^\star  \qquad .
 \end{align}
 Similarly the last Einstein equation $\mathcal E_{\tau\,\tau}=0$ leads to an equation for the function $\gamma_{\tau\,\tau}(\tau,\,\theta)$~:
\begin{align} \dot \gamma_{\tau\,\tau}(\tau,\,\theta)=&-Q^2\, \gamma''_{\tau\,\tau}(\tau,\,\theta)+a(\tau) \gamma'_{\tau\,\tau}(\tau,\,\theta)+2 \,L\, \omega\,   \dot a(\tau)\label{evolg}
\end{align}
whose solution reads :
 \begin{align}
 \gamma_{\tau\,\tau}(\tau,\,\theta) =&2\, \omega\,   a(\tau)+Q^2\,V_{(0,0)}(\tau,\,\theta)\\
\mbox{with}\ :\ V_{(0,0)}(\tau,\,\theta)=&\sum_n V^{(n)}_{(0,0)} \,e^{i \,n(A(\tau)+\theta)+n^2 \,Q^2\,\tau}\quad \mbox{and} \quad V^{(-n)}_{(0,0)}=(V^{(n)}_{(0,0)})^\star\qquad.
 \end{align}
Then the remaining two gravitational equations are identically satisfied.
\par\noindent To make an end let us notice that by redefining the angular variable as :
\begin{align}{\theta+ A(\tau)\mapsto \theta}\end{align}
the metric simplifies into :
\begin{align}
\hspace{-15mm}g^{(B)}_{\mu\nu}=
\left(
\begin{array}{ccc}
-4\,Q^2\,\rho^2+L\,Q^2\,\rho\, V_{(0,0)}(\tau,\,\theta)-\frac{k_g}4\,L^2\,Q^2\,m_{(0,0)}^{2} (\tau,\,\theta)&  -Q^2\,\rho\ &   L\, \omega\,  \rho \\
-Q^2\,\rho\  & 0  &   0\\
   L\,\omega\,  \rho &  0 &  \rho^2 
\end{array}
\right)\label{SpBS}
\end{align}
and, thanks to a residual gauge transformation, the abelian potential may be written as  : 
\begin{align}
A_\mu=\left\{\frac{L^2}\rho\,m_{(0,0)}(\tau,\,\theta),0,L\,c_\theta\right\}\qquad .\end{align}

\section{Surface charges}\label{AppSurfCh} 

\subsection*{A covariantisation lemma}
In this appendix we want to show that for second order diffeomorphism invariant theories, the surface charge density build by "cohomological methods'' (see for instance \cite{Barnich:2001jy,Compere:2018aar}) even if not explicitly covariant actually is covariant.\par\noindent
Let us denote tensorial background fields as $\phi_A$ and their variations as ${\delta \phi}_A$. 
The on-shell vanishing Noether current density can be expanded as :
\begin{align}
\underaccent{\cdot} s^\mu =\,& \sum_{n=0}^2  \underaccent{\cdot}  s^{\mu,(\nu)_n,B}\nabla_{(\nu)_n}{\delta \phi}_B = \sum_{n=0}^2 \tilde{\underaccent{\cdot} s}^{\mu,(\nu)_n,B}\partial_{(\nu)_n}{\delta \phi}_B \qquad ,\label{twoexp}
\end{align}
where all the coefficients $\underaccent{\cdot}  s^{\mu,(\nu)_n,B}$  are tensor densities, depending on the background fields and the reducibility parameters. 
The indices in the multisum are totally symmetrized. On-shell conservation $\partial_\mu \underaccent{\cdot} s^\mu = 0$ implies :
\begin{align}  {\underaccent{\cdot} s}^{(\mu,\nu_1\nu_2),B} =\,&0\qquad.\label{cons1}
\end{align}

The two expansion coefficients occurring in eq. \ref{twoexp} are related by :
\begin{align}
 \tilde{\underaccent{\cdot} s}^{\mu,\nu_1 \nu_2,B} =\,&  {\underaccent{\cdot} s}^{\mu,\nu_1 \nu_2,B}\qquad,\nonumber \\
 \tilde{\underaccent{\cdot} s}^{\mu,\nu,B}  =\,&  {\underaccent{\cdot} s}^{\mu,\nu ,B}- {\underaccent{\cdot} s}^{\mu,\nu_1 \nu_2,C} (\Gamma_{\nu_1\nu_2}^\nu \delta_C^B+ 2 \Gamma_{C(\nu_1}^B \delta_{\nu_2)}^\nu) \label{rel1}\qquad .
\end{align}
The not manifestly covariant expression for the surface charge density is given by : 
\begin{align}
\underaccent{\cdot}k^{\mu\nu}=\,& \tilde{\underaccent{\cdot} s}^{[\mu ,\nu],B} {\delta \phi}_B -\frac{2}{3} {\delta \phi}_B \partial_\rho  \tilde{\underaccent{\cdot} s}^{[\mu,\nu ] \rho,B} + \frac{4}{3}\partial_\rho {\delta \phi}_B  \tilde{\underaccent{\cdot} s}^{[\mu,\nu]\rho, B}\qquad ,
\end{align}
while the manifestly covariant one reads : 
\begin{align}
\underaccent{\cdot}\kappa^{\mu\nu}=\,&  {\underaccent{\cdot} s}^{[\mu ,\nu],B} {\delta \phi}_B -\frac{2}{3} {\delta \phi}_B \nabla_\rho  {\underaccent{\cdot} s}^{[\mu,\nu ] \rho,B} + \frac{4}{3}\nabla_\rho {\delta \phi}_B  {\underaccent{\cdot} s}^{[\mu,\nu]\rho, B}\qquad .
\end{align}
In order to check that on-shell $\underaccent{\cdot}k^{\mu\nu} = \underaccent{\cdot}\kappa^{\mu\nu}$, on one hand we expand $\underaccent{\cdot}k^{\mu\nu}$ using \eqref{rel1}, and on the other hand we expand the covariant derivatives in $\underaccent{\cdot}\kappa^{\mu\nu}$ and compare the two expressions.

The terms proportional to $\Gamma_{C \alpha}^B$ are given 
\begin{subequations}
\begin{align}
\text{in $\underaccent{\cdot}k^{\mu\nu}$ by :}\qquad  & -2 \tilde{\underaccent{\cdot} s}^{[\mu,\nu ] \nu_2,C}\Gamma_{C \nu_2}^B {\delta \phi}_B\qquad ,\\
\text{in $\underaccent{\cdot}\kappa^{\mu\nu}$ by :}\qquad   & \left( -\frac{2}{3} - \frac{4}{3} \right)   {\delta \phi}_B \tilde{\underaccent{\cdot} s}^{[\mu,\nu ] \rho,C} \Gamma_{C \rho}^B\qquad ,
\end{align}
\end{subequations}
and therefore are identical. \\
The terms proportional to $\Gamma_{\alpha\beta}^\rho$ are given 
\begin{subequations}
\begin{align}
\text{in $\underaccent{\cdot}k^{\mu\nu}$ by :}\qquad   & - {\underaccent{\cdot} s}^{[\mu, |\nu_1 \nu_2 | ,B } \Gamma_{\nu_1\nu_2}^{\nu ] } {\delta \phi}_B\qquad ,\\
\text{in $\underaccent{\cdot}\kappa^{\mu\nu}$ by :}\qquad  & - \frac{2}{3} {\delta \phi}_B \Gamma^{[\mu}_{\rho\nu_2} {\underaccent{\cdot} s}^{|\nu_2|, \nu ] \rho, B} - \frac{2}{3} {\delta \phi}_B \Gamma_{\rho \nu_2}^{[\nu} {\underaccent{\cdot} s}^{\mu],\nu_2 \rho,B} \\
=\,& + \frac{2}{3} {\delta \phi}_B \Gamma^{[\nu}_{\rho\nu_2} {\underaccent{\cdot} s}^{|\nu_2|, \mu ] \rho, B} - \frac{2}{3} {\delta \phi}_B \Gamma_{\rho \nu_2}^{[\nu} {\underaccent{\cdot} s}^{\mu],\nu_2 \rho,B} \qquad .
\end{align}
\end{subequations}
Therefore, their difference  is given by :
\begin{align}
 \,&  {\delta \phi}_B \Gamma^{[\nu}_{\nu_1 \nu_2} \left(- \frac{1}{3} {\underaccent{\cdot} s}^{\mu],\nu_1 \nu_2,B}  -\frac{2}{3}  {\underaccent{\cdot} s}^{|\nu_1|, \mu ] \nu_2, B} \right) \\
 =\,&- \frac{1}{2}  {\delta \phi}_B \Gamma^{\nu}_{\nu_1 \nu_2}  {\underaccent{\cdot} s}^{(\mu,\nu_1\nu_2),B} - (\mu \leftrightarrow \nu)\\
 =\,& 0 
\end{align}
in virtue of the conservation equation \eqref{cons1}.\\
Let us mention that this elementary proof does not straightforwardly generalises to higher order theories.

\subsection*{Explicit expressions}
In this appendix we write an explicit covariant expression of the surface charge density used in section (\ref{Charges}), based on the previous lemma. The diffeomorphism and gauge invariant action given by eq. (\ref{Action}) defines  the presymplectic potential  :
\begin{align}{\boldsymbol\Theta}^\mu=\,&\nabla_\alpha\delta g^{\mu\,\alpha}-\nabla^\mu\delta g-8\,\nabla^\mu\Phi\,\delta \Phi-\frac {k_g}4\,e^{-4\,\Phi}\,F^{\mu\,\alpha}\,\delta A_\alpha\nonumber\\
&-\frac 12\,e^{-8\,\Phi}\,H^{\mu\,\alpha\,\beta}\big(\delta B_{\alpha\beta}+\frac{k_g}2\,A_\alpha\,\delta A_\beta)\qquad ,\end{align}
where $\delta g_{\mu\nu}$, $\delta A_\alpha$ and $\delta B_{\alpha\beta}$ are arbitrary variation of the various fields ($\delta g^{\alpha\beta}:=g^{\alpha\,\mu}g^{\beta,\nu}\delta g_{\mu\nu}$, $\delta g=:g^{\mu\nu}\delta g_{\mu\nu}$).
Following standard techniques (i.e.  using the contracting homotopy operator; see for instance refs \cite{ Barnich:2001jy,Compere:2018aar}), we obtain the N\oe ther-Wald surface charge  :
\begin{align}Q^{\mu\nu}=\,&\frac 1{16\,\pi\,G}\Big(\nabla^\mu\,\xi^\nu+\nabla^\nu\,\xi^\mu+\frac {k_g}2\,e^{-4\,\Phi}F^{\mu\nu}(\xi^\alpha\,A_\alpha	+\,\lambda)\nonumber\\
&+e^{-8\,\Phi}H^{\mu\,\nu\,\alpha}\big(\xi^\beta\,B_{\beta\,\alpha }-\frac{k_g}4(\xi^\beta\,A_\beta+2\,\lambda)A_\alpha+\Lambda_\alpha\big)\Big)\qquad ,\label{NWsurfch}\end{align}
where $\{\zeta\}:=\{\xi^\mu,\,\Lambda_\alpha,\,\lambda\}$ are  gauge parameters. The surface charge density\footnote{We adopt here a commonly used terminology. Actually, what we are considering here is an "elementary surface charge density", an object that when integrated  will provide, in principle, a one-form on the phase space that still must be exact in order to allow to define true charges by an integration along a path, starting from a configuration where the values of the charges are prescribed by convention.} may be written as the sum of four contributions :
\begin{align} &\underaccent{\cdot}k^{\mu\nu}_{\{\zeta\}}=\frac {\sqrt{-g}}{16\,\pi\,G}\Big(k_E^{\mu\nu}+k_T^{\mu\nu}+k_H^{\mu\nu}+k_F^{\mu\nu}\Big)\label{NWk}\qquad .\end{align}
 A suffisant condition to be conserved ($\partial_\mu \underaccent{\cdot}k^{\mu\nu}_{\{\zeta\}}=0$) is that the background configuration is left invariant with respect to the gauge transformation generated by $\xi^\mu,\,\Lambda_\alpha$ and $\lambda$ (in which case they constitute a set of reducibility gauge parameters) and the field variations $\delta g_{\mu\nu}$, $\delta A_\alpha$ and $\delta B_{\alpha\beta}$ are solutions to the field equations linearised around this background.\\
Defining ${\mathfrak D}^{\sigma\mu\omega\nu}:=\,g^{\sigma\omega}g^{\mu\nu}-g^{\sigma\nu}g^{\mu\omega}$,  the general covariant expression of the surface charge is given by the sum of : 
\begin{itemize}
\item{\em Einstein tensor contribution}
 \begin{align}
 k^{ \mu\nu }_{E} =\,&\,2\left(\xi^{[\nu}\nabla^{\mu]}\delta g-\nabla_\rho \delta g^{\rho[\mu}\xi^{\nu]}+\xi^\rho\nabla^{[\nu}\delta g^{\mu]}_\rho\right)-\left(\nabla^{[\mu}\xi^{\nu]}\delta g-\nabla_\rho\xi^{[\nu}\,\delta g^{\mu]\rho}+\delta g_\rho^{[\mu}\nabla^{\nu]}\xi^\rho\right)
\end{align}
\item{\em Energy-momentum tensor contribution}
\begin{align}
k^{\mu\nu}_T=\,&\ 8\left(\xi^\nu\,\nabla^\mu\Phi-\xi^\mu\,\nabla^\nu\Phi\right)\delta f\nonumber \\
& -\frac 12 e^{-8\,\Phi}\,\xi^{[\mu}\,\eta^{\nu]\rho\sigma}\,\star H\left(\delta B_{\rho\sigma}+\frac {k_g}2\, A_{\rho}\,\delta A_{\sigma}\right){\nonumber  }\\
&+\frac {k_g} 2\,e^{-4\,\Phi}\left(\xi_{\sigma}\,F^{\sigma[\mu}\,g^{\nu]\rho}-3\,\xi^{[\rho}\,F^{\mu\nu]}\right) \delta A_\rho
\end{align}
Here $\star H:=\frac 1{3!}\eta^{\alpha\beta\gamma}H_{\alpha\beta\gamma}$  and the Schouten identity is used to simplify the second term.
\item{\em Scalar field contribution}
\par\noindent  As there is no derivative of the gauge parameter in case of  scalar fields, these fields only contribute via the other field equations 
\item{\em 2-form field contribution} ($\overline\lambda:=\xi^\rho A_{\rho}+\lambda$)
\begin{align}
k^{\mu\nu}_F=\,&\ \frac{k_g}2\left\{ 
\overline\lambda\,e^{-4\,\Phi}\,{{\mathfrak D}}^{\rho\sigma\nu\mu}\nabla_\rho \delta A_\sigma-\frac12\,{{\mathfrak D}}^{\rho\sigma\nu\mu}\, \nabla_\rho(\overline\lambda\,e^{-4\,\Phi})\,\delta A_\sigma\phantom{\frac{k_g}4}\nonumber \right.\nonumber \\
&\phantom{\ \frac{k_g}2\ } +\overline\lambda\, e^{-4\,\Phi}\left[\left(2\,g^{\rho[\nu}F^{\mu]\sigma}-\frac 12\,F^{\mu\nu}g^{\rho\sigma}\right)\delta g_{\rho\sigma}+2\,{{\mathfrak D}}^{\mu\nu\sigma\rho}(\nabla_\sigma\Phi)\delta A_\rho+4\,F^{\mu\nu}\,\delta f\right]\nonumber \\
&\phantom{\ \frac{k_g}2\ } \left.+\overline\lambda\, e^{-8\,\Phi}\left[g^{\rho[\nu}F^{\mu]\sigma}\,\delta B_{\rho\sigma}+ H^{\mu\nu\rho}\,\delta A_\rho+\frac{k_g}2\,g^{\rho[\mu}F^{\nu]\sigma}\,\delta A_{[\rho}\,A_{\sigma]}\right]\right\}
\end{align}

\item{\em 3-form field contribution} ($\overline \Lambda_\alpha:=\xi^\rho B_{\rho\alpha}-\frac{k_g}4(\xi^\rho A_\rho+2\,\lambda)\,A_\alpha +\Lambda_\alpha$)
\begin{align}
k^{\mu\nu}_H=\,&\,e^{-8\,\Phi}\left\{-\frac13\,\overline\Lambda_\beta\left(\eta^{\mu\beta(\nu}\eta^{\rho)\sigma\omega}-\eta^{\nu\beta(\mu}\eta^{\rho)\sigma\omega}\right)\left(\nabla_\rho\, \delta B_{\sigma\omega}-\frac{k_g}2\,A_\omega\,\nabla_{\rho}\,\delta A_\sigma\right)
\right.\nonumber \\
&\phantom{e^{-8\,\Phi}}+4\,\nabla_\omega \Phi\,\overline\Lambda_\beta\,\eta^{\beta\omega[\mu}\eta^{\nu] \rho\sigma}\,\left(\delta B_{\rho\sigma}-\frac{k_g}2\,A_\sigma\,\delta A_\rho\right)
\nonumber \\
&\phantom{e^{-8\,\Phi}\ }+\left.\frac{k_g}4\,\overline\Lambda_{\beta}\,\nabla_\sigma A_\omega\left(\eta^{\mu\beta[\nu}\eta^{\sigma]\omega\rho}-\eta^{\nu\beta[\mu}\eta^{\sigma]\omega\rho}\right) \delta A_\rho
+\frac12\,\overline\Lambda_\omega\,H^{\mu\nu\omega}\,h\delta g
+8\,\overline\Lambda_\beta\,H^{\mu\nu\beta}\delta f\right\}{\nonumber  }\\
&\phantom{e^{-8\,\Phi}\ }+\frac1{6}\,\nabla_\rho\left[e^{-8\,\Phi}\,\overline\Lambda_\beta\left(\eta^{\mu\beta(\nu}\eta^{\rho)\sigma\omega}-\eta^{\nu\beta(\mu}\eta^{\rho)\sigma\omega}\right)\right]\, \delta B_{\sigma\omega}
\nonumber \\
&\phantom{e^{-8\,\Phi}\ }-\frac {k_g}{12}\nabla_\rho\left[e^{-8\,\Phi}\,\overline\Lambda_\beta\,A_\omega\,\left(\eta^{\mu\beta(\nu}\eta^{\rho)\sigma\omega}-\eta^{\nu\beta(\mu}\eta^{\rho)\sigma\omega}\right)\right] \delta A_\sigma    
\end{align}
\end{itemize}
\subsection*{Non conserved charges}
The definition of physically relevant charges in the framework of gravity is a subtle question. Inserting the expressions of the asymptotic solutions obtained in sect. (\ref{IntasympEqs}) and asymptotic symmetries (\ref{xias}, \ref{BondiGaugeCond}) in the previous expression of the surface charge density, will in general leads to ill defined (divergent) expressions. Nevertheless we may {\it a priori} restrict the phase space by conditions insuring that the charges so obtained are finite, even if they remain in general dependent on the curves on which the charge density is integrated and thus, in particular, time dependent. The relevance of such time dependent charge is particularly well illustrated by the Bondi mass formula that describes the conversion of mass into gravitational radiation(see for instance \cite{Bondi:1962, Sachs:1962, burg:1966nr}). \\
Expanding the charge density (\ref{stark}) in terms of the radial coordinate $\rho$ we obtain :
\begin{align}
\star k=&\big(\rho\,k_{\tau(0,-1)}(\tau,\,\theta) +\ln^2[\rho/L]\,k_{\tau(2,0)}(\tau,\,\theta)+\ln[\rho/L]\,k_{\tau(1,0)}(\tau,\,\theta)\big)d\tau\nonumber\\
&+\big(\ln^2[\rho/L]\,k_{\theta(2,0)}(\tau,\,\theta)+\ln[\rho/L]\,k_{\theta(1,0)}(\tau,\,\theta)\big)d\theta\nonumber\\
&+\frac 1 \rho\big(\ln[\rho/L]\,k_{\rho(1,1)}(\tau,\,\theta)+k_{\rho(0,1)}(\tau,\,\theta)\big)d\rho\nonumber\\
&+\mathcal O[1]\qquad .
\end{align}
This charge density is defined up to an exact one-form $dF(\tau,\,\rho,\,\theta)$. Using this freedom we obtain the following conditions of finiteness.\\
\begin{itemize}
\item To avoid linear divergences :
\begin{subequations}
\begin{align}
  16\,\pi\,G\,k_{\tau(0,-1)}(\tau,\,\theta)=& \Big(\dot Y(\tau,\,\theta)-Q^2\,Y''(\tau,\,\theta)\Big)\frac{\delta Q}Q-\frac 14\,k_g\,Q^2\Big(\dot \ell(\tau,\,\theta)+c_\theta\,\dot Y(\tau,\,\theta)\Big)\delta c_\theta\nonumber\\
 =& 0
 \end{align}
 and some squared logarithmic ones :
 \begin{align}
 16\,\pi\,G\big(k_{\theta(2,0)}(\tau,\,\theta)-\frac 12 \partial_\theta k_{\rho(1,1)}(\tau,\,\theta)\big)=& -\frac 14 \Big(\ell'(\tau,\,\theta)+c_\theta\,Y'(\tau,\,\theta)\Big)\,\delta \Big(Q^2\,a^{(0)}_{\theta(1,1)}\Big)\nonumber\\
&=0\qquad .
\end{align}
\end{subequations}
From these two conditions we infer that :
\begin{align}
 Y(\tau,\,\theta)=\sum_n Y^{(n)}e^{i\,n\,\theta-n^2\,Q^2\,\tau}\qquad,\qquad
 \ell(\tau,\,\theta)=-c_\theta\,Y(\tau,\,\theta)+\lambda_c \qquad .
 \end{align}
 Let us emphasise the asymptotic behaviour of $Y(\tau,\,\theta)$ that goes to a constant when $\tau$ goes to infinity, on the contrary to the various solutions of the heat equation met in the main text.
\item There remains a squared logarithmic divergent term.  Inserting in its expression  the previous conditions, we obtain :
 \begin{align}
16\,\pi\,G\big(k_{\tau(2,0)}(\tau,\,\theta)-\frac 12 \partial_\tau k_{\rho(1,1)}(\tau,\,\theta)\big)=&  -2\,L\,\dot X(\tau)\,\delta \Big(Q^2\,f'_{(0,1)}(\tau,\,\theta)\Big)\nonumber\\
&-L\,X(\tau)\,\delta\Big(Q^4\,f_{(0,1)}'''(\tau,\,\theta)\Big)\nonumber\\
 =&0
 \end{align}
which imposes for being satisfied that we restrict $f_{(0,1)}(\tau,\,\theta)$ to be a constant $f_{(0,1)}^{(0)}$ and as a consequence the constant $f_{(1,1)}^{(0)} $  to be zero (see eqs (\ref{Jacf01}, \ref{Jacf11}).\\
\item Finally, to avoid logarithmic divergences two more conditions have to be satisfied :
{\small
 { \begin{subequations}
 \begin{align}
 16\,\pi\,G\Big(k_{\theta(1,0)}(\tau,\,\theta)- \partial_\theta k_{\rho(0,1)}(\tau,\,\theta)\Big)= &k_g\,L\,Q^2\Big(X(\tau)\big(2\,a_{(1,1)}^{(0)}(\tau,\,\theta)
  -\frac 12\,Q^2\,a''_{(0,1)}(\tau,\,\theta)\big)\nonumber\\
 &\phantom{k_g\,L\,Q^2 }-\frac 12\big(Y'(\tau,\,\theta)\,a^{(0)}_{(1,1)}-Y(\tau,\,\theta)\,a'_{(0,1)}(\tau,\,\theta)\big)\Big)\delta c_\theta\nonumber\\
 &-4\frac LQ\,Y'(\tau,\,\theta)\,\delta \Big(Q\,f_{(0,1)}^{(0)}\Big)\nonumber\\
=&0\\
16\,\pi\,G\Big(k_{\tau(1,0)}(\tau,\,\theta)-  \partial_\tau k_{\rho(0,1)}(\tau,\,\theta)\Big)=&L\,\dot X(\tau)\,\delta\Big(Q^2 \,n_{(0,1)}(\tau,\,\theta)\Big)-4\,L\,Q\,Y''(\tau,\,\theta)\,\delta\Big(Q\,f_{(0,1)}^{(0)}\Big)\nonumber\\
&+\frac 14  L\Big(2\,X(\tau)\,n''_{(0,1)}(\tau,\,\theta)-k_g \,a''_{(0,1)}(\tau,\,\theta)\,\lambda_c\Big) \,\delta\Big(Q^4\Big)\nonumber\\
&-\frac 14 \,k_g\,L\,Q^4\Big(\big(Y(\tau,\,\theta)\,a'_{(0,1)}(\tau,\,\theta)\big)'
+\big(X(\tau)\,a'_{(0,1)}(\tau,\,\theta)\big )\dot\strut\nonumber\\
&\phantom{-\frac 14 \,k_g\,L\,Q^4}+Y''(\tau,\,\theta)\,a^{(0)}_{(1,1)}\Big)\,\delta c_\theta\nonumber\\
&= 0 
\end{align}
\end{subequations}}}
\end{itemize}

We shall not pursue the discussion of the various restrictions that may be imposed on the phase space to satisfy these conditions. The important point  to notice is that subdominant terms of the asymptotic expansions contribute to the divergent part of the asymptotic charges.
\subsection*{A comment about asymptotic charges and diffeomorphisms}
Practically, on 3-dimensional  spaces as such considered in the main text,  the elementary charges at infinity are given by integrals on closed loops located at infinity~:
\begin{align}{{\mathchar'26\mkern-9mu \delta}}{\mathcal Q}=\int_{{\mathcal C}} \star k_\mu\,dx^\mu\label{elemch}\end{align}
of a one-form $ \star k=k_\mu\,dx^\mu$ whose differential goes to zero as well as its integral on two-dimensional surfaces (denoted $\mathcal S$)  located near infinity. Here by infinity, we refer to the region parametrised by a radial variable ($\rho$ or $r$) that grows without limit.\\
More precisely, on the ${\tau, \rho,\theta}$ patch the one form 
$\star k$ reads
\begin{align}\star k=k_\tau(\tau,\,\rho,\,\theta)\,d\tau+k_\rho(\tau,\,\rho,\,\theta)\,d\rho+k_\theta(\tau,\,\rho,\,\theta)\,d\theta\qquad .\label{stark}\end{align}
Let us consider the loops given by the circles :
\begin{align}\overline{{\mathcal C}}[\tau_0,\,\rho_0]=\,&\{\tau=\tau_0,\,\rho=\rho_0,\,\theta\in[0,2\,\pi]\}\qquad .\label{Cbar}\end{align} 
Elementary charges are given by the integrals :
\begin{align} {\mathchar'26\mkern-9mu \delta} \overline{\mathcal Q}=\,&\lim_{\rho_0\rightarrow\infty}\int_{\overline{\mathcal C}[\tau_0,\,\rho_0]} \star k_\mu\,dx^\mu=\lim_{\rho_0\rightarrow\infty}\int_0
^{2\,\pi} k_\theta(\tau_0,\rho_0,\theta)d\theta\qquad .\end{align}
Performing a diffeomorphism like those considered in the main text [eqs (\ref{diffeoTH}, \ref{TebHQF})] :
\begin{align}\tau=T(u)=\,&\int_0^ue^{b0(\upsilon)}\,d\upsilon\quad ,\quad  \rho=r/(Q\,\partial_\phi F(u,\phi))=r\,e^{-2\,f_{(0,0)}(u,\,\phi)}/Q\quad ,{\nonumber  }\\
\theta =\,&Q\,F(u,\,\phi) =Q\Big(\int_0^\phi e^{2\,f_{(0,0)}(u,\,\xi)}\,d\xi+h(u)\Big)\quad ,\label{diffeo}\end{align}
 the one-form $\star k$ transforms covariantly and reads :
 \begin{align}\star k=k_u(u,\,r,\,\phi)\,du+k_r(u,\,r,\,\phi)\,dr+k_\phi(u,\,r,\,\phi)\,d\phi\qquad .\end{align}
 
 On the new patch so defined, let us assume that elementary charges ${{\mathchar'26\mkern-9mu \delta}} \tilde{\mathcal Q}$ are obtained from integrations on the circles~: 
 \begin{align}\tilde{\mathcal C}[u_0,r_0]=\,&\{u=u_0,\,r=r_0,\phi\in[0,2\,\pi]\}\end{align}
 in the limit $r_0\rightarrow \infty$ :
 \begin{align}{{\mathchar'26\mkern-9mu \delta}} \tilde{\mathcal Q}=\,&\lim_{r_0\rightarrow\infty}\int_0
^{2\,\pi} k_\phi(u_0,r_0,\phi)d\phi\label{dQti}\\
=\,&\lim_{r_0\rightarrow\infty}\int_0^{2\,\pi}(\tilde{k}_\theta\,e^{2\,f_{(0,0)}(u_0,\,\phi)}-\tilde{k}_\rho\,r\,e^{-2\,f_{(0,0)}(u_0,\,\phi)}\,\partial_\phi f_{(0,0)}(u_0,\,,\phi)\big)d\phi\\
\mbox{where }&\nonumber 
\tilde{k}_{\theta,(resp.\ \rho)}(u,\,r,\,\phi) =k_{\theta,(resp.\ \rho)} \big(T(u ),\ft r Q e^{-2\,f_{(0,0)}(u ,\phi)},Q\,F(u ,\phi)\big)\qquad.\end{align} 
  These circles does not coincide with the previous ones [eq.(\ref{Cbar})] but nevertheless ${{\mathchar'26\mkern-9mu \delta}} \overline{\mathcal Q}$ will be equal to ${{\mathchar'26\mkern-9mu \delta}} \tilde{\mathcal Q}$ if asymptotically $d\star k$ and its integral on  asymptotic surfaces $\mathcal S$ such that their boundaries are $\partial \mathcal S=\tilde C-\overline C$ also vanishes :
 \begin{align}\lim_{r_0\rightarrow \infty}\int_{\mathcal S}d\star k=0\qquad\mbox{implies that}\qquad {{\mathchar'26\mkern-9mu \delta}} \tilde{\mathcal Q}={{\mathchar'26\mkern-9mu \delta}} \overline{\mathcal Q}\qquad .\end{align}
 Moreover in such case the elementary charges are conserved.
 
 However, we also have to take into account that the gauge parameters being covariantly transformed the two charges so obtained, while numerically equal didn't refer to the same physical quantities\footnote{I'm indebeted to Marc Henneaux, for a clarifying discussion on this point.}. Indeed  let us suppose that
we are considering,  using coordinates $\{x^\alpha\}:=\{\tau,\,\rho,\,\theta\}$, an elementary charge with respect to the vector field :
\begin{align}\xi^\alpha=\{X(\tau),\,-\rho\,\partial_\theta Y(\tau,\,\theta) , \, Y(\tau,\,\theta)\}\qquad .\label{genxi}\end{align} Once having performed the coordinate change defined by eqs (\ref{diffeo}), that maps $\{x^\alpha\}$ on $\{y^\beta\}:=\{u,\,r,\,\phi\}$, the charge given by eq. (\ref{dQti}) refers to another configuration, physically distinct, of the system (the previous one transformed by a large diffeomorphism) 
but with respect the another vector  of components $\zeta^\beta$ obtained by acting with the Jacobian $\partial y^\beta/\partial x^\alpha$ on $\xi^\alpha$ : $$\zeta^u=1/\partial_u T(u)\,X(T(u))\ , \dots,$$ $$ \zeta^\phi=(e^{-2\,f_{(0,0)}(u,\,\phi)}/Q)\xi^\theta-(e^{-2\,f_{(0,0)}(u,\,\phi)}/\partial_u T(u))\partial_\phi F(u,\,\phi)\,\xi^\tau\quad .$$ Thus the obtainment, for a black string configuration with an arbitrary asymptotic expression $f(u,\,\phi)$ of the dilaton field, of the charge associated to an asymptotic Killing vector $\xi$ such that $\xi^\phi=e^{i\,n\,\phi}$ require the knowledge of the inverse of the transformation eqs (\ref{diffeo}). A task that in general is difficult to do analytically.\\
But as in the context of the field configurations discussed in the main text the charges provided by general $\xi^\alpha$ (\ref{genxi}) is not conserved, and becomes dependent (in order to be finite) on the choice of the asymptotic curves on which the density is integrated, we shall not pursue their discussion.
\section*{Acknowledgments}

This work found its inspiration from several sources. The main one was a proposal made by S. Detournay, more than two years ago, to study the asymptotic charges and  thermodynamics of the black string configurations we obtained some years ago \cite{Detournay:2005fz}. Elements of answer to this problem were developed with him, M. Petropoulos and C. Zwikel. Needless to say that I warmly acknowledge all three for discussing my considerations  and developments on the problem.

In particular I want also to thanks M. Petropoulos  for having explained me in detail the relation between the special solution I obtained and an extra deformation of the $SL(2,{\mathbb R})$ {\sc{wzw}} model, using para-fermions operators and S. Detournay for informing me that this work presents an overlap with a forthcoming article \cite{Detournay:2018cbf}.
There also are others colleagues and friends that I also want to acknowledge for various enlightening discussions:  G. Barnich, D. Bini, N. Boulanger, G. Comp\`ere, P. Chru\'sciel, T.~Damour, R. Garani, G. Guillermin, M. Henneaux, S. Massar, U. Moschella, B. Olbak and, last but not least, IHES (France) for its kind hospitality.

\end{document}